%% file: main.tex
\documentclass{article}

\usepackage[utf8]{inputenc}
\usepackage[T1]{fontenc}

\PassOptionsToPackage{table, dvipsnames}{xcolor}
\usepackage{xcolor}

\PassOptionsToPackage{colorlinks=true, citecolor=blue, linkcolor=blue, urlcolor=black}{hyperref}

\usepackage{arxiv}
\usepackage[utf8]{inputenc}
\usepackage[T1]{fontenc}
\usepackage{url}
\usepackage{booktabs}
\usepackage{nicefrac}
\usepackage{microtype}
\usepackage{lipsum}
\usepackage{graphicx}
\usepackage{natbib}
\usepackage{doi}
\usepackage{algorithm}
\usepackage{multirow}
\usepackage{multicol}
\usepackage{amsmath,amssymb,amsfonts}
\usepackage{amsthm}
\usepackage{mathrsfs}
\usepackage{float}
\usepackage{algorithmic}
\usepackage{bm}
\usepackage{caption}
\usepackage{hyperref}

\title{Inhomogeneous plane waves in \\ attenuative anisotropic porous media
}

\author{
	{Lingli~Gao, ~Weijian~Mao, ~Qianru~Xu, ~Wei~Ouyang, ~Shaokang~Yang} \\
	Research Center for Computational and Exploration Geophysics, \\State Key Laboratory of Precision Geodesy,\\
    Innovation Academy for Precision Measurement Science and Technology, \\ Chinese Academy of Sciences, Wuhan 430077, China \\
        \And
	\href{https://orcid.org/0000-0001-8868-7967}{\includegraphics[scale=0.06]{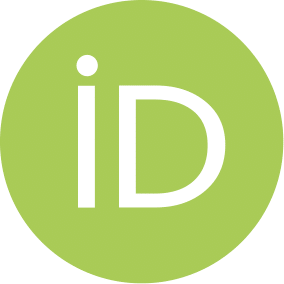}\hspace{1mm}Shijun~Cheng} \\
	Division of Physical Science and Engineering\\
	King Abdullah University of Science and Technology\\
	Thuwal 23955-6900, Saudi Arabia \\
    [3ex]
  $^{*}$Corresponding author: \textbf{Shijun Cheng}~(\texttt{sjcheng.academic@gmail.com})
}

\renewcommand{\shorttitle}{Inhomogeneous plane waves in attenuative anisotropic porous media}

\hypersetup{
pdftitle={A template for the arxiv style},
pdfsubject={q-bio.NC, q-bio.QM},
pdfauthor={David S.~Hippocampus, Elias D.~Striatum},
pdfkeywords={First keyword, Second keyword, More},
}

\begin{document}
\maketitle

\input{Sections/Abstract}
\input{Sections/Introduction}
\input{Sections/Method}
\input{Sections/Numerical_Examples}

\input{Sections/Discussion}
\input{Sections/Conclusions}

\input{Sections/CodeAvailability}
\input{Sections/Appendix}

\bibliographystyle{unsrtnat}
\bibliography{references}

\end{document}

%% file: Sections/Abstract.tex
\begin{abstract}
Seismic wave propagation in poro-viscoelastic anisotropic media is of practical importance for exploration geophysics and global seismology. Existing theories generally utilize homogeneous plane wave theory, which considers only velocity anisotropy but neglects attenuation anisotropy and wave inhomogeneity arising from attenuation. As a result, it poses significant challenges to accurately analyzing seismic wave dispersion and attenuation in poro-viscoelastic anisotropic media. In this paper, we investigate the propagation of inhomogeneous plane waves in poro-viscoelastic media, explicitly incorporating both velocity and attenuation anisotropy. Starting from classical Biot theory, we present a fractional differential equation describing wave propagation in attenuative anisotropic porous media that accommodates arbitrary anisotropy in both velocity and attenuation. Then, instead of relying on the traditional complex wave vector approach, we derive new Christoffel and energy balance equations for general inhomogeneous waves by employing an alternative formulation based on the complex slowness vector. The phase velocities and complex slownesses of inhomogeneous fast and slow quasi-compressional (qP1 and qP2) and quasi-shear (qS1 and qS2) waves are determined by solving an eighth-degree algebraic equation. By invoking the derived energy balance equation along with the computed complex slowness, we present explicit and concise expressions for energy velocities. Additionally, we analyze dissipation factors defined by two alternative measures: the ratio of average dissipated energy density to either average strain energy density or average stored energy density. We clarify and discuss the implications of these definitional differences in the context of general poro-viscoelastic anisotropic media. Finally, our expressions are degenerated to give their counterparts of the homogeneous waves as a special case, and the reduced forms are identical to those presented by the existing poro-viscoelastic theory. Several examples are provided to illustrate the propagation characteristics of inhomogeneous plane waves in unbounded attenuative vertical transversely isotropic porous media.
\end{abstract}

\keywords{Seismic anisotropy \and Seismic attenuation \and Wave propagation \and Elasticity and anelasticity \and Fracture and flow}

%% file: Sections/Introduction.tex
\section{\textbf{Introduction}}
It is widely known that the recorded seismic waves carry the property information of the Earth's interior. Therefore, providing an accurate wave propagation theory is very important to understand wave propagation in real formation. This is especially true in the field of studying the structure of the Earth, predicting the microscopic properties of rocks, and seismic exploration. In the past decades, the topics of wave propagation in complex media have attracted considerable attention \citep{robertsson1994viscoelastic, zhu2014theory}. Numerous studies indicate that most of the Earth’s interior exhibits seismic anisotropy, which may be caused by a series of factors, such as material microstructure, composition, properties, compaction, and the existence of aligned fractures and microcracks \citep{zhu2014theory, zhu2017numerical}. For convenience, \cite{crampin1984seismic} classified materials with some form of seismic anisotropy into three categories: intrinsic anisotropy, fracture induced anisotropy, and long-wavelength anisotropy. When a seismic wave propagates in specific materials, it will present velocity and attenuation anisotropy, that is, both velocity and attenuation are directionally dependent. Most existing literature is devoted to focusing on velocity anisotropy, e.g., vertical transverse isotropy (VTI) \citep{thomsen1986weak, jin2019reflection}, tilted transverse isotropy (TTI) \citep{grechka2001velocity, pvsenvcik2017reflection, mu2020modeling}, orthorhombic anisotropy (OA) \citep{tsvankin1997anisotropic, stovas2018geometric, stovas2023singularity, stovas2024degenerate}, and general anisotropy \citep{vavryvcuk1999weak, stovas2024singularity}. Actually, laboratory and field measurements of rock samples suggest that the attenuation anisotropy is sometimes more significant than the velocity anisotropy \citep{arts1992approximation, prasad2003velocity, carter2006attenuation, vcerveny2008quality, usher2017measuring}. Hence, the study of wave behavior in the presence of attenuation anisotropy is essential to further understand the reservoir properties.

The early research on attenuation anisotropy mainly used Backus’s theory and mechanical models to simulate anisotropy-$Q$ in viscoelastic finely layer under long-wavelength approximation \citep{carcione1992anisotropic, picotti2010q, picotti2012oscillatory}. However, there is a maximum effective frequency limit for using the long-wavelength approximation, which limits the frequency range of interest. Alternatively, some researchers extended the relaxation time of the mechanical model to the anisotropic relaxation time, thereby realizing the seismic modeling with arbitrary attenuation anisotropy in a broad range of frequencies \citep{komatitsch1999introduction, bai2016time}. The huge computational and memory consumption caused by introducing the anisotropic relaxation time hinders large-scale practical applications. \cite{zhu2017numerical} presented that the anisotropic $Q$ matrix is incorporated into the relaxation function of the constant-$Q$ model \citep{kjartansson1979constant} to obtain a viscoelastic anisotropic wave equation that enables to model the anisotropic attenuation, which is solved by Grünwald-Letnikov approximation in the time domain. To solve a large amount of the time history of field variables storage caused by time-domain simulation, \cite{zhu2019efficient} suggested replacing fractional time derivatives with fractional Laplacians and using the Fourier pseudospectral method to implement the viscoelastic anisotropic wave equation numerically. \cite{wang2022propagating} derived a decoupled fractional Laplacian viscoelastic wave equation that captures frequency‐independent $Q$ in VTI media and demonstrate its superior accuracy for heterogeneous $Q$ models, directional attenuation modeling, and computational efficiency. Although the above wave propagation model can be employed to better simulate seismic wave propagation in a single-phase medium, a viscoelastic anisotropic porous solid saturated with a viscous fluid are closer to the actual media beneath the surface. Therefore, the study of seismic wave propagation in poro-viscoelastic anisotropic media has more reasonable significance. 

Only less seismological literatures have studied the seismic waves propagation in poro-viscoelastic anisotropic media. Currently, poro-viscoelasticity mainly uses the viscoelastic model to express the relaxed skeleton (RS) or the relaxed fluid diffusion (RF) mechanisms in the porous media. In the related research papers \citep{cheng2021wave, yang2021wave}, the RS mechanism is used to describe the wave-induced relative vibration between the porous solids, that is, the relaxation relationship between the stress and strain of the solid skeleton. The RF mechanism implies that the friction force, generated by the relative motion between the solid skeleton and the fluid, is not simply proportional to the velocity of the fluid relative to the solid, but a relaxation relationship varied with time \citep{carcione1996wave, carcione2013theory}. The RS and RF mechanisms act on the seismic wave dispersion and attenuation in the seismic exploration bands ($<10^3$ Hz) and the high-frequency bands ($>10^3$ Hz), respectively. \cite{ba2008skeleton} unified these two mechanisms into the poro-viscoelastic model to realize the modeling of wave behavior in a wide frequency band, but they did not consider viscous fluid pressure. It’s emphasized that most of these research papers are restricted to velocity anisotropy, and attenuation is assumed to be isotropic. However, attenuation anisotropy may affect the wave behavior more significantly, especially in cases of fluid flow has a dominant direction. Thus, a more realistic poro-viscoelastic model should present a wave equation that can describe the complete anisotropy (i.e., velocity and attenuation anisotropy). Recently, \cite{han2023incorporating, han2025wave} explicitly incorporated both velocity and attenuation anisotropy into poro-viscoelastic models: one study embeds first- and second-order nearly constant-$Q$ relaxation functions into the Biot and Biot-squirt (BISQ) frameworks, while the other couples the BISQ equations with Kjartansson’s constant-$Q$ law via a sum-of-exponentials approximation. Both formulations are validated through finite-difference time-domain simulations of seismic wave propagation. However, neither work quantitatively assesses how attenuation anisotropy affects wave dispersion and attenuation, nor do they analyze these effects from a plane-wave perspective.

Actually, seismic waves propagation in poro-viscoelastic anisotropic media shows complicated behaviors. Due to poro-viscoelastic media are generally strongly dissipative media, seismic waves become inhomogeneous waves as they travel across poro-viscoelastic media, that is, the wave propagation direction is not parallel to the attenuation direction \citep{carcione2014wave}. In this case, a plane wave analysis based on the assumption of a homogeneous wave may not be able to explain the attenuation beyond a certain level. Therefore, a more general analysis should start with inhomogeneous waves. Much attention has been devoted to the propagation characteristics of inhomogeneous plane waves in viscoelastic single-phase media \citep{carcione1993energy, vcerveny2005plane1, vcerveny2005plane2, vcerveny2006energy, vcerveny2008quality, behura2009role}. However, the studies of inhomogeneous plane waves propagating in poro-viscoelastic media are not so common. \cite{carcione2014wave} presented the pioneering studies, using the method of the wave vector to provide some basic relations of inhomogeneous and homogeneous plane waves in poro-viscoelastic media. However, he did not derive the Christoffel equation and the expressions of energy velocity, phase velocity, and dissipation factor for inhomogeneous waves. Based on the complex slowness vector, \cite{sharma2010energy} derived the energy balance equation for a poro-viscoelastic anisotropic solid saturated with viscous fluid. Also, he did not present explicit expressions of measurable quantities. \cite{liu2020seismic} studied the dissipation factor of inhomogeneous waves in effective Biot theory, but for isotropic porous media. 

In this work, a poro-viscoelastic anisotropic model is upscaled from the classical Biot theory \citep{biot1956theory1, biot1956theory2, biot1962generalized, biot1962mechanics}, which includes both RS and RF mechanisms and enables to model the complete anisotropy. Here, differ from \cite{carcione1996wave}, we not only include RF mechanism in dynamic Darcy's law, but also consider the effects of RF mechanism in fluid pressure and constitutive relation. The attenuation anisotropy of solid skeleton and fluid flow is characterized by introducing the matrices $\boldsymbol{Q}^s$ and $\boldsymbol{Q}^f$, respectively, in which the components of the two anisotropic $\boldsymbol{Q}$ matrices are used to represent the fractional order of time in the relaxation function of RS and RF mechanisms, respectively. The mathematical formulations of relaxation functions, which are expressed as a power law and contain a gamma function \citep{caputo1971new}, are the same as that of constant-$Q$ wave propagation theory \citep{kjartansson1979constant}. Using the property of the gamma function, the time convolution in the dynamic Darcy’s law, constitutive relation, and the fluid pressure can be replaced by the time-fractional derivatives.

To comprehensively understand the properties of inhomogeneous plane waves propagation in this new poro-viscoelastic anisotropic model, we exactly present the expressions of the measurable quantities (i.e., phase and energy velocities, dissipation factor) and the basic relations (i.e., Christoffel equation, energy balance equation, and the relationship between phase and energy velocities). We use the method based on mixed specification of the complex slowness vector \citep{vcerveny2005plane1, vcerveny2005plane2, vcerveny2006energy} to derive the Christoffel equation. The equation has a similar form as that of homogeneous plane waves provided by \cite{carcione2014wave} using the complex wave vector. However, there are substantial differences between the matrices and vectors terms in the equations. The complex slowness and phase velocities of the inhomogeneous fast and slow quasi-compressional (qP1 and qP2) and quasi-shear (qS1 and qS2) waves are determined by solving an eighth-degree algebraic equation, whereas the corresponding complex polarization vector is calculated by substituting the complex slowness into the derived Christoffel equation. The energy balance equation, which describes the dynamic process of seismic wave propagation, is a prerequisite for deriving energy velocities and dissipation factors. \cite{carcione2014wave} presents the complex form of energy balance equation in poro-viscoelastic media. However, considering the attenuation anisotropy, we convert the variation of fluid content into a vector, which results in extended stress and strain vectors having a different form from the Carcione definition. For this reason, the equations proposed by Carcione cannot be applied here. Therefore, we deduce a new energy balance equation for attenuative anisotropic porous media. The difference between our and Carcione's equations rests with the calculation of average strain and dissipated strain energy densities. 

Based on the presented energy balance equation, the energy velocities and dissipation factors of inhomogeneous waves are derived and can be explicitly expressed in terms of slowness vector, wave inhomogeneity, and the known material parameters. Here, the dissipation factors are derived by invoking different definitions $Q_1^{-1}$ (i.e., the ratio of the average dissipated energy density to twice the average strain energy density; see \cite{carcione2014wave} and  $Q_2^{-1}$ (i.e., the average dissipated energy density divided by the average stored energy density; see \cite{buchen1971plane}). The reason for this is that these two classical definitions are quite different in the case of inhomogeneous waves, which has been verified by \cite{liu2020seismic} in effective isotropic Biot materials. We here clarify the difference between the two definitions in more general poro-viscoelastic anisotropic materials. Also, we verify that the relationship between energy and phase velocities still holds in attenuative anisotropic porous media, that is, the phase velocity is the projection of energy velocity in the propagation direction, which has been reported in poro-viscoelastic anisotropic \citep{carcione2014wave} and single-phase viscoelastic anisotropic media \citep{carcione1993energy}. We degenerate the derived explicit expressions to the special case of homogeneous waves, and discuss the differences between the proposed theory and previous poro-viscoelastic models.

To illustrate the proposed theory, we consider a porous medium with VTI attenuation commonly used to describe shales or other layered rocks, and numerically study the wave behavior characteristics of inhomogeneous waves in 2-D unbounded attenuative VTI porous media, especially the analysis of dispersion and attenuation. The derived concise and explicit expressions of energy velocities and dissipation factors are used for this purpose. We first analysis the proposed model, that is, the model can show the behavior of attenuation anisotropy and explain a stronger dispersion and attenuation in a broad range of frequencies by comparing it with other poro-viscoelastic models. Then, we illustrate the effects of attenuation anisotropy and inhomogeneity on the behavior of the qP1, quasi-shear (qS), and qP2 waves, and compare the difference of dissipation factors under the definitions  $Q_1^{-1}$ and $Q_2^{-1}$. 

%% file: Sections/Method.tex
\section{\textbf{Theory}}\label{sec:theory}
In this section, we develop a comprehensive theoretical framework to study seismic wave propagation in attenuative anisotropic porous media. Initially, we revisit classical Biot’s poroelastic theory to describe wave-induced fluid flow (WIFF). Recognizing the limitations of Biot models in accurately capturing wave attenuation and dispersion, we introduce a generalized poro-viscoelastic model, which incorporates both velocity and attenuation anisotropy by embedding fractional viscoelastic mechanisms into the constitutive relations and the dynamic Darcy’s law. Subsequently, the model is utilized to derive the Christoffel equation explicitly tailored for inhomogeneous plane wave propagation using a complex slowness vector formalism. Additionally, we provide explicit expressions for measurable physical quantities, including energy velocities and dissipation factors, for general inhomogeneous waves based on energy balance considerations. Finally, our derived equations are demonstrated to reduce to their established counterparts for homogeneous wave conditions, ensuring alignment with existing theories.

From now on, the following annotations are used for convenience: the indices $i,j=1,2,3$ represent the spatial variables $x$, $y$, and $z$, respectively; $\partial _{i}^{m}$ and $\partial _{t}^{m}$ are the $m$-order spatial and time partial derivatives with respect to the variables ${{x}_{i}}$ and time $t$, respectively; the matrix notation is expressed by the upper case indices $I$ and $J$; the superscripts ‘$\operatorname{T}$’ and ‘$\nabla \cdot $’ denote the matrix transposition and divergence operator, respectively; the horizontal line above the variables represent the complex conjugate.

\subsection{Biot’s Theory}\label{sec:theorysub1}
Biot theory \citep{biot1956theory1, biot1956theory2, biot1962generalized, biot1962mechanics} claims that seismic wave propagation in the porous media can induce the pore fluid pressure gradient, which leads to fluid flow relative to the solid skeleton. In this process, due to the viscosity of the fluid, the energy of the seismic wave is lost by friction. Under the effective stress principle, the effective stress acting on the solid skeleton and the hydrostatic stress acting on the fluid constitutes the total stress $\boldsymbol{\sigma}$ acting on a porous medium. Therefore, for an anisotropic poroelastic medium saturated with a viscous fluid, the constitutive relationship can be expressed as \citep{biot1957elastic, thompson1991reformation}:
\begin{equation}\tag{1}\label{eq1} 
\boldsymbol{\sigma}=\boldsymbol{C}\cdot \boldsymbol{\varepsilon}-\boldsymbol{\alpha}P,
\end{equation} 
where $\boldsymbol{C}$ is the drained stiffness matrix, $\boldsymbol{\varepsilon}$ is the strain vector of the solid skeleton, $\boldsymbol{\alpha}$ is the Biot effective coefficient vector, and $P$ is the fluid pressure. The vectors $\boldsymbol{\sigma}$, $\boldsymbol{\varepsilon}$, and $\boldsymbol{\alpha}$ have the form:
\begin{equation}\tag{2}\label{eq2} 
\boldsymbol{\sigma }={{\left[ {{\sigma }_{11}},{{\sigma }_{22}},{{\sigma }_{33}},{{\sigma }_{23}},{{\sigma }_{13}},{{\sigma }_{12}} \right]}^{\operatorname{T}}}, 
\end{equation} 
\begin{equation}\tag{3}\label{eq3} 
\boldsymbol{\varepsilon }={{\left[ {{\varepsilon }_{11}},{{\varepsilon }_{22}},{{\varepsilon }_{33}},2{{\varepsilon }_{23}},2{{\varepsilon }_{13}},2{{\varepsilon }_{12}} \right]}^{\operatorname{T}}},
\end{equation} 
\begin{equation}\tag{4}\label{eq4} 
\boldsymbol{\alpha }={{\left[ {{\alpha }_{1}},{{\alpha }_{2}},{{\alpha }_{3}},{{\alpha }_{4}},{{\alpha }_{5}},{{\alpha }_{6}} \right]}^{\operatorname{T}}},
\end{equation}
where ${{\sigma }_{ij}}$ and ${{\varepsilon }_{ij}}$ are the stress and strain components, respectively, ${{\alpha }_{I}}\left( I=1,\cdot \cdot \cdot ,6 \right)$ are the Biot effective coefficients. In the most general case, drained stiffness matrix $\boldsymbol{C}$ has 21 components ${{C}_{IJ}}\left( I,J=1,\cdot \cdot \cdot ,6 \right)$. When the anisotropy of solid skeleton in porous media is not caused by the anisotropy of the individual grains but rather arises from their directional alignment, the Biot effective coefficients can be represented in terms of the elastic stiffness components \citep{yang2002poroelastic}.

In Biot’s theory, the fluid pressure for the poroelastic anisotropic media can be expressed as \citep{biot1962mechanics}
\begin{equation}\tag{5}\label{eq5} 
P=M\left( \zeta -{{\boldsymbol{\alpha }}^{\text{T}}}\cdot \boldsymbol{\varepsilon} \right),
\end{equation} 
where $\zeta =-\nabla \cdot \boldsymbol{w}$ denotes the variation of fluid content, $\boldsymbol{w}=\varphi \left( \boldsymbol{U}-\boldsymbol{u} \right)$ is the relative fluid displacement multiplied by material porosity $\varphi$, $\boldsymbol{U}=\left( {{U}_{1}},{{U}_{2}},{{U}_{3}} \right)$ and $\boldsymbol{u}=\left( {{u}_{1}},{{u}_{2}},{{u}_{3}} \right)$ represent the absolute fluid and solid displacements, respectively, and the modulus $M$ is given by
\begin{equation}\tag{6}\label{eq6} 
M=\frac{K_{s}^{2}}{{{K}_{s}}\left[ 1+\varphi \left( {{K}_{s}}K_{f}^{-1}-1 \right) \right]-{C_{ava}}\;},
\end{equation} 
where ${{K}_{f}}$ and ${{K}_{s}}$ are the bulk modulus of the fluid and solid grains, respectively, and $C_{ava}=\left( {{C}_{11}}+{{C}_{22}}+{{C}_{33}}+2{{C}_{12}}+2{{C}_{13}}+2{{C}_{23}} \right)/9$.

The dynamic equations of wave propagation in porous media are given by \citep{biot1962generalized}
\begin{equation}\tag{7}\label{eq7}
\sum\limits_{j=1}^{3}{{{\partial }_{j}}{{\sigma }_{ij}}}+{{f}_{i}}=\rho {{\partial }_{t}}{{v}_{i}}+{{\rho }_{f}}{{\partial }_{t}}{{q}_{i}},
\end{equation} 
where ${{f}_{i}}$ are the components of the body forces vector, ${{v}_{i}}={{\partial }_{t}}{{u}_{i}}$ and ${{q}_{i}}=\varphi {{\partial }_{t}}({{U}_{i}}-{{u}_{i}})$ are the particle velocities of the solid and fluid (relative to the solid), respectively, $\rho =\left( 1-\varphi  \right){{\rho }_{s}}+\varphi {{\rho }_{f}}$ is the composite density, ${{\rho }_{s}}$ and ${{\rho }_{f}}$ denote the solid and fluid densities, respectively. 

The dynamic motion of WIFF relative to the solid skeleton is described by the dynamic Darcy's law as follows \citep{biot1962generalized}
\begin{equation}\tag{8}\label{eq8}
-{{\partial }_{i}}P={{\rho }_{f}}{{\partial }_{t}}{{v}_{i}}+Y({{p}^{B}}){{q}_{i}},
\end{equation} 
where $Y({{p}^{B}})$ is the Biot viscodynamic operator and is a function of time. The operator $Y({{p}^{B}})$ embodies the dynamics of the fluid in relative motion, which depends on the geometry of the pores, fluid inertia, and viscosity. Several expressions have been developed for the viscodynamic operator \citep{biot1962generalized, johnson1987theory}, such as the expression of the dynamic Darcy's law proposed by \cite{carcione1998viscoelastic} as follows
\begin{equation}\tag{9}\label{eq9}
-{{\partial }_{i}}P={{\rho }_{f}}{{\partial }_{t}}{{v}_{i}}+{{\phi }_{i}}{{\partial }_{t}}{{q}_{i}}+\frac{\eta }{{{K}_{ii}}}{{q}_{i}},    \end{equation} 
where ${{\phi }_{i}}={{{T}_{i}}{{\rho }_{f}}}/{\varphi }\;$, with the tortuosity ${{T}_{i}}$, $\eta $ is the fluid viscosity, and ${{K}_{ii}}$ are the components of the third-rank rank permeability matrix $\boldsymbol{K}$. The permeability $\boldsymbol{K}$ is closely related to the geometric characteristics of the solid skeleton, that is, the anisotropic solid skeleton results in anisotropic permeability. For example, in a poroelastic VTI medium, ${{K}_{11}}$ and ${{K}_{22}}$ are usually equal and ${{K}_{12}}={{K}_{13}}={{K}_{23}}=0$.

Biot theory is the most classic theory for studying the propagation of elastic waves in fluid-saturated porous media, laying the foundation for the wave theory of porous media. However, it cannot be ignored that the Biot model is still significantly lower in predicting the dispersion and attenuation of seismic waves, especially in the seismic exploration band \citep{liu2018effective, zhang2021wave}. Therefore, in the following, we present a poro-viscoelastic model to make up for the deficiencies of the Biot model.

\subsection{Attenuative anisotropic porous media model}\label{sec:theorysub2}
The introduction of viscoelasticity can reasonably explain the strong dispersion and attenuation in reservoir rocks, which has been verified in the researches of many poro-viscoelastic models \citep{ba2008skeleton, cheng2021wave}. Most researches on the poro-viscoelastic model are restricted to velocity anisotropy, while ignoring the attenuation anisotropy, which is unreasonable. Here, we incorporate both velocity and attenuation anisotropy in the poro-viscoelastic model and, thus, propose an attenuative anisotropic porous medium model based on the Biot theory.

Unlike a single-phase medium, which only involves the attenuation anisotropy of solid skeleton, a porous medium under the framework of Biot theory is a two-phase medium composed of solid skeleton and pore fluid. As stated earlier, the pore fluid flow is closely related to the solid skeleton. This signifies that the attenuation anisotropy needs to be considered in both the solid skeleton and the fluid. Thus, a more reasonable attenuative anisotropic porous medium model should include anisotropic RS and RF mechanisms. In this case, in terms of the constitutive relationship (\ref{eq1}) of the classic Biot theory, the time-domain constitutive relationship for the attenuative anisotropic porous media model should be expressed as:
\begin{equation}\tag{10}\label{eq10}
\boldsymbol{\sigma }\text{=}{{\boldsymbol{\psi }}^{\left( 1 \right)}}*{{\partial }_{t}}\boldsymbol{\varepsilon }-{{\boldsymbol{\psi }}^{\left( 2 \right)}}*{{\partial }_{t}}P,
\end{equation}
where the symbol $*$ denotes the time convolution, ${{\boldsymbol{\psi }}^{\left( 1 \right)}}=\left[ \psi _{IJ}^{\left( 1 \right)} \right]$ and ${{\boldsymbol{\psi }}^{\left( 2 \right)}}={{\left( \psi _{1}^{\left( 2 \right)},\psi _{2}^{\left( 2 \right)},\psi _{3}^{\left( 2 \right)},\psi _{4}^{\left( 2 \right)},\psi _{5}^{\left( 2 \right)},\psi _{6}^{\left( 2 \right)} \right)}^{\text{T}}}$ are relaxation function matrix and vector, respectively. ${{\boldsymbol{\psi }}^{\left( 1 \right)}}$ and ${{\boldsymbol{\psi }}^{\left( 2 \right)}}$ correspond to the relaxation relationship between the total stress $\boldsymbol{\sigma }$ of porous media and skeleton strain $\boldsymbol{\varepsilon }$ and fluid pressure $P$, respectively, which are caused by the RS and RF mechanisms, respectively.

Meanwhile, since the fluid diffusion mechanism is a relaxation relationship that changes with time, the fluid pressure is no longer non-viscous, but needs to be regarded as viscous. Therefore, the fluid pressure equation (\ref{eq5}) needs to be recast as
\begin{equation}\tag{11}\label{eq11}
P={{\boldsymbol{\psi }}^{\left( 3 \right)}}*{{\partial }_{t}}\boldsymbol{\zeta }-{{\boldsymbol{\psi }}^{\left( 4 \right)}}*{{\partial }_{t}}\boldsymbol{\varepsilon },    \end{equation} 
where ${{\boldsymbol{\psi }}^{\left( 3 \right)}}=\left( \psi _{1}^{\left( 3 \right)},\psi _{2}^{\left( 3 \right)},\psi _{3}^{\left( 3 \right)} \right)$ and ${{\boldsymbol{\psi }}^{\left( 4 \right)}}=\left( \psi _{1}^{\left( 4 \right)},\psi _{2}^{\left( 4 \right)},\psi _{3}^{\left( 4 \right)},\psi _{4}^{\left( 4 \right)},\psi _{5}^{\left( 4 \right)},\psi _{6}^{\left( 4 \right)} \right)$ are relaxation function vectors, which control the variation of fluid pressure and are assumed to be determined only by the RF mechanism. It should be noted that here we turn the quantity $\zeta $ into a vector form $\boldsymbol{\zeta }={{\left( {{\zeta }_{1}},{{\zeta }_{2}},{{\zeta }_{3}} \right)}^{\text{T}}}$, where the components ${{\zeta }_{i}}=-{{\partial }_{i}}{{w}_{i}}$, considering that the attenuation anisotropy causes the differential variation of fluid content in the $x$, $y$, and $z$ directions.

Furthermore, the influence of the RF mechanism should also be reflected in the dynamic Darcy’s law. The classic Biot theory holds that energy dissipation depends only on the wave-induced relative motion between the fluid and the solid. The diffusion force is generated by the relative motion, which is proportional to the fluid velocity relative to the solid. When considering the RF mechanism, the diffusion force becomes a relaxation relationship. Therefore, we modify the dynamic Darcy’s law (\ref{eq9}) as follows:
\begin{equation}\tag{12}\label{eq12}
-{{\partial }_{i}}P={{\rho }_{f}}{{\partial }_{t}}{{v}_{i}}+{{\phi }_{i}}{{\partial }_{t}}{{q}_{i}}+\psi _{i}^{\left( 5 \right)}*{{\partial }_{t}}{{q}_{i}},
\end{equation}
where $\psi _{i}^{\left( 5 \right)}$ are the components of the relaxation function vector ${{\boldsymbol{\psi }}^{\left( 5 \right)}}=\left( \psi _{1}^{\left( 5 \right)},\psi _{2}^{\left( 5 \right)},\psi _{3}^{\left( 5 \right)} \right)$. This way of considering the RF mechanism in porous media has been developed in many literatures \citep{carcione1996wave, carcione2013theory}. Here, we also use this method because it can provide the effect of dispersion and attenuation in the high-frequency range. 

The matrix ${{\boldsymbol{\psi }}^{\left( 1 \right)}}$ and the vectors ${{\boldsymbol{\psi }}^{\left( 2 \right)}}$, ${{\boldsymbol{\psi }}^{\left( 3 \right)}}$, ${{\boldsymbol{\psi }}^{\left( 4 \right)}}$, and ${{\boldsymbol{\psi }}^{\left( 5 \right)}}$ contain all the wave behavior information of attenuative anisotropic porous medium under infinitesimal deformations and are usually represented by the mechanical and the fractional viscoelastic models. Compared with the mechanical models, the fractional viscoelastic model may be more suitable for describing the anelastic behavior in porous media \citep{yang2021wave}. It can also be used to approximate the dispersion and attenuation caused by WIFF \citep{picotti2017numerical}. Hence, we refer the frequency-independent fractional $Q$ model \citep{kjartansson1979constant} to represent the quantities ${{\boldsymbol{\psi }}^{\left( 1 \right)}}$, ${{\boldsymbol{\psi }}^{\left( 2 \right)}}$, ${{\boldsymbol{\psi }}^{\left( 3 \right)}}$, ${{\boldsymbol{\psi }}^{\left( 4 \right)}}$, and ${{\boldsymbol{\psi }}^{\left( 5 \right)}}$. 

The mathematical formulations of the frequency-independent fractional $Q$ model are \citep{kjartansson1979constant}
\begin{equation}\tag{13.1}\label{eq13.1}
\sigma = \psi * \partial_{t}\varepsilon
          = M_{0}\,\omega_{0}^{-2\gamma}\,\partial_{t}^{2\gamma}\varepsilon,
\end{equation}
\begin{equation}\tag{13.2}\label{eq13.2}
\psi = \frac{M_{0}}{\Gamma\bigl(1-2\gamma\bigr)}
       \Bigl(\frac{t}{t_{0}}\Bigr)^{-2\gamma}H(t),
\end{equation}
where $\sigma$ and $\varepsilon$ represent the stress and strain, respectively, $\psi $ is the relaxation function, ${{M}_{0}}=\rho {{v}_{0}}^{2}{{\cos }^{2}}(\pi \gamma /2)$ is a bulk modulus, $\rho$ is density, ${{v}_{0}}$ is the phase velocity at the reference frequency ${{\omega }_{0}}$, ${{t}_{0}}={1}/{{{\omega }_{0}}}\;$ is a reference time, $\gamma ={\arctan \left( {1}/{Q}\; \right)}/{\pi }\;$ is a dimensionless parameter in the range of $\left[ 0,0.5 \right]$ for any positive value of quality factor $Q$, $\Gamma $ denotes Euler’s gamma function, and $H\left( t \right)$ is the Heaviside step function. Using the relaxation function (\ref{eq13.2}) expressed as a power law, the convolution can be replaced by the time-fractional derivative \citep{zhu2017numerical}. The time-fractional derivative is determined by the past value, that is, the stress in the current state needs to be calculated from the time history of strain. This can be well applied to explain the RS and RF mechanisms in porous media. For example, the relaxation function (\ref{eq13.2}) is introduced into the modified Darcy's law, that is, considering the time-fractional derivative, which enables to describe the variation of permeability during the process of diffusion. This may occur because the fluid may carry solid particles that block the pores or may chemically react with the solid skeleton.

Therefore, analogous to the equation (\ref{eq13.2}), we define the components $\psi _{IJ}^{\left( 1 \right)}$ of the matrix ${{\boldsymbol{\psi }}^{\left( 1 \right)}}$ as
\begin{equation}\tag{14}\label{eq14}
\psi _{IJ}^{\left( 1 \right)}=\frac{{{C}_{IJ}}{{\cos }^{2}}\left( \pi {{\gamma }_{IJ}}/2 \right)}{\Gamma \left( 1-2{{\gamma }_{IJ}} \right)}{{\left( \frac{t}{{{t}_{0}}} \right)}^{-2{{\gamma }_{IJ}}}}H\left( t \right),
\end{equation}
where the index $\left(\textbf{x},t \right)$ ($\textbf{x}$ is the position vector) of $\psi _{IJ}^{\left( 1 \right)}$ has been dropped for convenience, ${{C}_{IJ}}$ are the drained stiffness coefficients, and the anisotropic ${{\gamma }_{IJ}}={\arctan \left( {1}/{Q_{IJ}^{s}}\; \right)}/{\pi }\;$ can be expressed by introducing an anisotropic matrix ${{\boldsymbol{Q}}^{s}}$ that represents the RS mechanism of a general attenuation anisotropy as follows:
\begin{equation}\tag{15}\label{eq15}
{{\boldsymbol{Q}}^{s}}=\left[ \begin{matrix}
   Q_{11}^{s} & Q_{12}^{s} & Q_{13}^{s} & Q_{14}^{s} & Q_{15}^{s} & Q_{16}^{s}  \\
   Q_{12}^{s} & Q_{22}^{s} & Q_{23}^{s} & Q_{24}^{s} & Q_{25}^{s} & Q_{26}^{s}  \\
   Q_{13}^{s} & Q_{23}^{s} & Q_{33}^{s} & Q_{34}^{s} & Q_{35}^{s} & Q_{36}^{s}  \\
   Q_{14}^{s} & Q_{24}^{s} & Q_{34}^{s} & Q_{44}^{s} & Q_{45}^{s} & Q_{46}^{s}  \\
   Q_{15}^{s} & Q_{25}^{s} & Q_{35}^{s} & Q_{45}^{s} & Q_{55}^{s} & Q_{56}^{s}  \\
   Q_{16}^{s} & Q_{26}^{s} & Q_{36}^{s} & Q_{46}^{s} & Q_{56}^{s} & Q_{66}^{s}  \\
\end{matrix} \right]. 
\end{equation}
Here, we follow Carcione's definition of anisotropic $Q$ \citep{carcione2014wave}, that is, $Q_{IJ}^{s}$ are the ratios of the real and the imaginary parts of the corresponding complex stiffness coefficients. This means that the anisotropic matrix ${{\boldsymbol{Q}}^{s}}$ inherits the structure of the stiffness matrix $\boldsymbol{C}$, resulting in the same structure of the matrices $\boldsymbol{C}$ and ${{\boldsymbol{Q}}^{s}}$. Thus, nine independent parameters ($Q_{11}^{s}$, $Q_{12}^{s}$, $Q_{13}^{s}$, $Q_{22}^{s}$, $Q_{23}^{s}$, $Q_{33}^{s}$, $Q_{44}^{s}$, $Q_{55}^{s}$, and $Q_{66}^{s}$) are needed to describe the orthorhombic anisotropic attenuation of solid skeleton. For porous media with TI attenuation, five independent parameters ($Q_{11}^{s}$, $Q_{13}^{s}$, $Q_{33}^{s}$, $Q_{55}^{s}$, and $Q_{66}^{s}$) are required to characterize the anelastic behavior. If the attenuation is considered to be isotropic, the matrix ${{\boldsymbol{Q}}^{s}}$ only needs independent parameters $Q_{33}^{s}$ and $Q_{55}^{s}$, where $Q_{13}^{s}=Q_{12}^{s}=Q_{23}^{s}$ is calculated by $Q_{13}^{s}=Q_{33}^{s}\left( {{C}_{33}}-2{{C}_{55}} \right){{\left( {{C}_{33}}-{2{{C}_{55}}Q_{33}^{s}}/{Q_{55}^{s}}\; \right)}^{-1}}$ as proposed in the single-phase media with an attenuative anisotropic solid skeleton by \cite{zhu2006plane}. 

The vectors ${{\boldsymbol{\psi }}^{\left( 2 \right)}}$, ${{\boldsymbol{\psi }}^{\left( 3 \right)}}$, ${{\boldsymbol{\psi }}^{\left( 4 \right)}}$, and ${{\boldsymbol{\psi }}^{\left( 5 \right)}}$ are only related to the RF mechanism and, therefore, we define them as
\begin{equation}\tag{16.1}\label{eq16.1}
\psi _{I}^{(2)}=\frac{{{\alpha }_{I}}{{\cos }^{2}}\left( \pi {{\nu }_{I}}/2 \right)}{\Gamma \left( 1-2{{\nu }_{I}} \right)}{{\left( \frac{t}{{{t}_{0}}} \right)}^{-2{{\nu }_{I}}}}H\left( t \right)~~(I=1,\cdot \cdot \cdot ,6),
\end{equation}

\begin{equation}\tag{16.2}\label{eq16.2}
\psi _{I}^{(3)}=\frac{M{{\cos }^{2}}\left( \pi {{\nu }_{I}}/2 \right)}{\Gamma \left( 1-2{{\nu }_{I}} \right)}{{\left( \frac{t}{{{t}_{0}}} \right)}^{-2{{\nu }_{I}}}}H\left( t \right)~~(I=1,2,3),       \end{equation}

\begin{equation}\tag{16.3}\label{eq16.3}
\psi _{I}^{(4)}=\frac{M{{\alpha }_{I}}{{\cos }^{2}}\left( \pi {{\nu }_{I}}/2 \right)}{\Gamma \left( 1-2{{\nu }_{I}} \right)}{{\left( \frac{t}{{{t}_{0}}} \right)}^{-2{{\nu }_{I}}}}H\left( t \right) ~~(I=1,\cdot \cdot \cdot ,6),
\end{equation}

\begin{equation}\tag{16.4}\label{eq16.4}
\psi _{I}^{(5)}=\frac{\eta {{\cos }^{2}}\left( \pi {{\nu }_{I}}/2 \right)}{{{K}_{II}}\Gamma \left( 1-2{{\nu }_{I}} \right)}{{\left( \frac{t}{{{t}_{0}}} \right)}^{-2{{\nu }_{I}}}}H\left( t \right)~~(I=1,2,3).
\end{equation}

Similarly, in equations (\ref{eq16.1})-(\ref{eq16.4}), the parameters ${{\nu }_{I}}={\arctan \left( {1}/{Q_{I}^{f}}\; \right)}/{\pi }\;$ can be calculated from the anisotropic matrix ${{\boldsymbol{Q}}^{f}}$ characterizing the RF mechanism of attenuation anisotropy as follows:
\begin{equation}\tag{17}\label{eq17}
{{\boldsymbol{Q}}^{f}}=\left[ \begin{matrix}
   Q_{1}^{f} & Q_{6}^{f} & Q_{5}^{f}  \\
   Q_{6}^{f} & Q_{2}^{f} & Q_{4}^{f}  \\
   Q_{5}^{f} & Q_{4}^{f} & Q_{3}^{f}  \\
\end{matrix} \right].
\end{equation}
We assume that the attenuation anisotropic symmetry of the RF mechanism is the same as that of the RS mechanism. Hence, if the solid skeleton behaves the behavior of orthorhombic anisotropic attenuation, the matrix ${{\boldsymbol{Q}}^{f}}$ has three independent parameters ($Q_{1}^{f}$, $Q_{2}^{f}$, $Q_{3}^{f}$). When the solid skeleton has the TI attenuation, the matrix ${{\boldsymbol{Q}}^{f}}$ is described by two independent parameters. Let us consider an attenuation isotropic, in which the matrix ${{\boldsymbol{Q}}^{f}}$ only requires one parameter. 

Substituting each relaxation function into the corresponding equations, and replacing the convolution with the time-fractional derivative yields
\begin{equation}\tag{18.1}\label{eq18.1}
\boldsymbol{\sigma }\text{=}{{\boldsymbol{C}}^{0}}\cdot \boldsymbol{\varepsilon }-{{\boldsymbol{\alpha }}^{0}}P,
\end{equation}
\begin{equation}\tag{18.2}\label{eq18.2}
P=M\left[ {{\left( {{\boldsymbol{\alpha }}^{1}} \right)}^{\text{T}}}\cdot \boldsymbol{\zeta }-{{\left( {{\boldsymbol{\alpha }}^{0}} \right)}^{\text{T}}}\cdot \boldsymbol{\varepsilon } \right],
\end{equation}
\begin{equation}\tag{18.3}\label{eq18.3}
-{{\partial }_{i}}P={{\rho }_{f}}{{\partial }_{t}}{{v}_{i}}+{{\phi }_{i}}{{\partial }_{t}}{{q}_{i}}+b_{ii}^{0}\partial _{t}^{2{{\nu }_{i}}}{{q}_{i}},
\end{equation}
where the components of the matrix ${{\boldsymbol{C}}^{0}}=\left[ C_{IJ}^{0} \right]$ and the vectors ${{\boldsymbol{\alpha }}^{0}}={{\left( \alpha _{1}^{0},\alpha _{2}^{0},\alpha _{3}^{0},\alpha _{4}^{0},\alpha _{5}^{0},\alpha _{6}^{0} \right)}^{\text{T}}}$ and ${{\boldsymbol{\alpha }}^{1}}={{\left( \alpha _{1}^{1},\alpha _{2}^{1},\alpha _{3}^{1} \right)}^{\text{T}}}$ are given as
\begin{equation}\tag{19.1}\label{eq19.1}
C_{IJ}^{0}={{C}_{IJ}}{{\cos }^{2}}(\pi {{\gamma }_{IJ}}/2)\omega _{0}^{-2{{\gamma }_{IJ}}}\partial _{t}^{2{{\gamma }_{IJ}}}~~(I,J=1,\cdot \cdot \cdot ,6),
\end{equation}
\begin{equation}\tag{19.2}\label{eq19.2}
\alpha _{I}^{0}={{\alpha }_{I}}{{\cos }^{2}}(\pi {{\nu }_{I}}/2)\omega _{0}^{-2{{\nu }_{I}}}\partial _{t}^{2{{\nu }_{I}}}~~(I=1,\cdot \cdot \cdot ,6),
\end{equation}
\begin{equation}\tag{19.3}\label{eq19.3}
\alpha _{I}^{1}={{\cos }^{2}}(\pi {{\nu }_{I}}/2)\omega _{0}^{-2{{\nu }_{I}}}\partial _{t}^{2{{\nu }_{I}}}~~(I=1,2,3),
\end{equation}
and the coefficients $b_{ii}^{0}$ are
\begin{equation}\tag{19.4}\label{eq19.4}
b_{ii}^{0}=\frac{\eta }{{{K}_{ii}}}{{\cos }^{2}}\left( \pi {{\nu }_{i}}/2 \right)\omega _{0}^{-2{{\nu }_{i}}}~~(i=1,2,3).
\end{equation}
Here, the terms ${{\nu }_{i}}$ $(i=1,2,3)$ and ${{\nu }_{I}}$ $(I=1,2,3)$ are equivalent.

From equations (\ref{eq19.1})-(\ref{eq19.4}), we can see that: 
\begin{enumerate}
    \item When the components $Q_{IJ}^{s}$ approach infinity, the matrix ${{\boldsymbol{C}}^{0}}$ approach the matrix $\boldsymbol{C}$, and the attenuative anisotropic porous media degenerates into a poro-viscoelastic media with the RF mechanism;
    \item If the components $Q_{I}^{f}$ are close to infinity, the parameters ${{\nu }_{I}}$ are nearly equal to zero, which causes the attenuative anisotropic porous media to become a poro-viscoelastic media considering only the RS mechanism;
    \item In the case where both are infinite, the attenuative anisotropic porous media model turns into a poroelastic media under the framework of the classic Biot theory.
\end{enumerate}

The existing poro-viscoelastic models only present the velocity anisotropy but ignore the attenuation anisotropy. Meanwhile, the analysis of the relaxation mechanism in the porous medium is incomplete, and the relaxation relationship is only introduced in the solid skeleton (i.e., the part $\boldsymbol{C}\cdot \boldsymbol{\varepsilon }$ in equation (\ref{eq1})) and the dynamic Darcy’s law (i.e., equation (\ref{eq12})). In contrast, the proposed model includes both attenuation anisotropy in the fluid and the solid skeleton, and a complete relaxation mechanism is fairly considered in the constitutive relations, dynamic Darcy’s law, and fluid pressure (see equations (\ref{eq18.1})-(\ref{eq18.3})), which is closer to the properties of actual media. Although more relaxation functions are introduced, they only need to be represented by the components of two anisotropic matrices ${{\boldsymbol{Q}}^{s}}$ and ${{\boldsymbol{Q}}^{f}}$ at the reference frequency ${{\omega }_{0}}$. This signifies that we only add a few parameters to describe the wave behavior in complex attenuative anisotropic porous media.

\subsection{Inhomogeneous plane waves and the Christoffel equation in attenuative anisotropic porous media}\label{sec:theorysub3}
Although \cite{carcione2014wave} presented the Christoffel equation in poro-viscoelastic anisotropic media, the derivation is started from homogeneous plane wave propagation, considering only RF mechanism and attenuation isotropic. Seismic waves propagation in an attenuative porous medium are generally inhomogeneous waves. That is, the propagation direction of seismic waves is different from the maximum attenuation direction \citep{liu2020seismic}. If the seismic wave is assumed to propagate as a plane wave, there is an included angle between the real and the imaginary parts of the slowness vector, which means that the planes of constant amplitude and constant phase do not coincide \citep{zhu2006plane}. Hence, in the following, without loss of generality, we consider the inhomogeneous plane wave propagation in 3-D attenuative anisotropic porous media, and the Christoffel equation is derived by invoking the inhomogeneous plane wave analysis method based on the complex slowness vector.

Considering a time-harmonic field $\exp \left( -\text{i}\omega t \right)$, where $\text{i}=\sqrt{-1}$, a new expression relates to the extended stress vector $\boldsymbol{T}$ to the extended strain vector $\boldsymbol{\Pi }$ as follows:
\begin{equation}\tag{20}\label{eq20}
\boldsymbol{T}=\boldsymbol{R}\cdot \boldsymbol{\Pi}, 
\end{equation}
where $\boldsymbol{T}$ and $\boldsymbol{\Pi }$ are defined as
\begin{equation}\tag{21.1}\label{eq21.1}
{{\boldsymbol{T}}^{\operatorname{T}}}=\left[ {{\sigma }_{11}},{{\sigma }_{22}},{{\sigma }_{33}},{{\sigma }_{23}},{{\sigma }_{13}},{{\sigma }_{12}},-P,-P,-P \right],           \end{equation}
\begin{equation}\tag{21.2}\label{eq21.2}
{{\boldsymbol{\Pi }}^{\operatorname{T}}}=\left[ {{\varepsilon }_{11}},{{\varepsilon }_{22}},{{\varepsilon }_{33}},2{{\varepsilon }_{23}},2{{\varepsilon }_{13}},2{{\varepsilon }_{12}},-{{\zeta }_{1}},-{{\zeta }_{2}},-{{\zeta }_{3}} \right],
\end{equation}
and $\boldsymbol{R}$ is a complex matrix with the components as follows:
\begin{equation}\tag{22}\label{eq22}
R_{IJ} \;=\;
\begin{cases}
C_{IJ}^*+M\alpha_I^*\alpha_J^*,
&1\le I,J\le 6,\\[1ex]
{M\alpha_I^*\alpha_{J-6}^*}/{\alpha_{J-6}},
&1\le I\le6,\;7\le J\le9,\\[1ex]
{M\alpha_J^*},
&7\le I\le9,\;1\le J\le6,\\[1ex]
{M\alpha_{I-6}^*}/
      {\alpha_{I-6}},
&7\le I,J\le9.
\end{cases}
\end{equation}
where
\begin{equation}\tag{23.1}\label{eq23.1}
C_{IJ}^{*}={{C}_{IJ}}{{\cos }^{2}}(\pi {{\gamma }_{IJ}}/2)\omega _{0}^{-2{{\gamma }_{IJ}}}{{\left( -\text{i}\omega  \right)}^{2{{\gamma }_{IJ}}}}~~(I,J=1,\cdot \cdot \cdot ,6),       \end{equation}
\begin{equation}\tag{23.2}\label{eq23.2}
\alpha _{I}^{*}={{\alpha }_{I}}{{\cos }^{2}}(\pi {{\nu }_{I}}/2)\omega _{0}^{-2{{\nu }_{I}}}{{\left( -\text{i}\omega  \right)}^{2{{\nu }_{I}}}}~~(I=1,\cdot \cdot \cdot ,6).
\end{equation}

The time derivative of the extended strain vector $\boldsymbol{\Pi }$ can be expressed in terms of the extended particle velocity vector $\boldsymbol{V}={{\left[ {{v}_{1}},{{v}_{2}},{{v}_{3}},{{q}_{1}},{{q}_{2}},{{q}_{3}} \right]}^{T}}$ as follows:
\begin{equation}\tag{24}\label{eq24}
{{\partial }_{t}}\boldsymbol{\Pi }={{\boldsymbol{D}}^{\operatorname{T}}}\cdot \boldsymbol{V},
\end{equation}
where $\boldsymbol{D}$ is a matrix composed of spatial partial differential operators, and can be expressed as
\begin{equation}\tag{25}\label{eq25}
\boldsymbol{D}=\left[ \begin{matrix}
   {{\partial }_{1}} & 0 & 0 & 0 & {{\partial }_{3}} & {{\partial }_{2}} & 0 & 0 & 0  \\
   0 & {{\partial }_{2}} & 0 & {{\partial }_{3}} & 0 & {{\partial }_{1}} & 0 & 0 & 0  \\
   0 & 0 & {{\partial }_{3}} & {{\partial }_{2}} & {{\partial }_{1}} & 0 & 0 & 0 & 0  \\
   0 & 0 & 0 & 0 & 0 & 0 & {{\partial }_{1}} & 0 & 0  \\
   0 & 0 & 0 & 0 & 0 & 0 & 0 & {{\partial }_{2}} & 0  \\
   0 & 0 & 0 & 0 & 0 & 0 & 0 & 0 & {{\partial }_{3}}  \\
\end{matrix} \right].
\end{equation}

Considering the time-harmonic field again, and using a convenient matrix notation, equations (\ref{eq24}) can be rewritten as
\begin{equation}\tag{26}\label{eq26}
-\operatorname{i}\omega \boldsymbol{\Pi }={{\boldsymbol{D}}^{\operatorname{T}}}\cdot \boldsymbol{V}, 
\end{equation}
and equations (\ref{eq7}) and (\ref{eq18.3}) become
\begin{equation}\tag{27}\label{eq27}
\boldsymbol{D}\cdot \boldsymbol{T}=-\text{i}\omega \boldsymbol{G}\cdot \boldsymbol{V},
\end{equation}
respectively, where
\begin{equation}\tag{28}\label{eq28}
\boldsymbol{G}=\left[ \begin{matrix}
   \rho  & 0 & 0 & {{\rho }_{f}} & 0 & 0  \\
   0 & \rho  & 0 & 0 & {{\rho }_{f}} & 0  \\
   0 & 0 & \rho  & 0 & 0 & {{\rho }_{f}}  \\
   {{\rho }_{f}} & 0 & 0 & {g_1} & 0 & 0  \\
   0 & {{\rho }_{f}} & 0 & 0 & {g_2} & 0  \\
   0 & 0 & {{\rho }_{f}} & 0 & 0 & {g_3}  \\
\end{matrix} \right]
\end{equation}
is a symmetric complex density matrix with $g_i={\phi }_{i}+b_{ii}^{0}{\left( -\text{i}\omega  \right)}^{2{\nu }_{i}-1}$, $i=1,2,3$. Combining equations (\ref{eq20}) and (\ref{eq26}) yields
\begin{equation}\tag{29}\label{eq29}
-\text{i}\omega \boldsymbol{T}=\boldsymbol{R}\cdot ({{\boldsymbol{D}}^{\operatorname{T}}}\cdot \boldsymbol{V}).
\end{equation}

The extended particle velocity vector $\boldsymbol{V}$ of inhomogeneous plane waves in the attenuative anisotropic porous material can be represented as \citep{liu2020seismic}
\begin{equation}\tag{30}\label{eq30}
\boldsymbol{V}={{\boldsymbol{V}}^{0}}\exp \left[ \text{i}\omega (\boldsymbol{s}\cdot \textbf{x}-t) \right],
\end{equation}
where ${{\boldsymbol{V}}^{0}}$ is a complex polarization vector, $\boldsymbol{s}$ is the complex slowness vector. The complex slowness vector can be expressed by various specifications, such as the directional, componental, and mixed specifications, which are discussed in detail by \cite{vcerveny2005plane1, vcerveny2005plane2} in a single-phase viscoelastic anisotropic medium. Among them, the mixed specification has stronger universality. Therefore, we refer the mixed specification to express the complex slowness vector in the attenuative porous medium as
\begin{equation}\tag{31}\label{eq31}
\boldsymbol{s}=s\boldsymbol{\hat{l}}=\tau \boldsymbol{\hat{n}}+\text{i}\kappa \boldsymbol{\hat{m}},
\end{equation}
where $s$ is the complex slowness, $\tau $ is an unknown complex quantity, $\kappa $ is a real-valued quantity, $\boldsymbol{\hat{l}}={{\left( {{l}_{1}},{{l}_{2}},{{l}_{3}} \right)}^{\text{T}}}$ is a complex unit vector along the direction of the complex slowness vector, $\boldsymbol{\hat{n}}={{\left( {{n}_{1}},{{n}_{2}},{{n}_{3}} \right)}^{\text{T}}}$ and $\boldsymbol{\hat{m}}={{\left( {{m}_{1}},{{m}_{2}},{{m}_{3}} \right)}^{\text{T}}}$ are the real-valued unit vectors perpendicular to each other ($\boldsymbol{\hat{n}}\cdot \boldsymbol{\hat{m}}=0$). Here, the vector $\boldsymbol{\hat{n}}$ represents the directions of wave propagation. The parameter $\kappa $, whose range is $\left[ -\infty ,+\infty  \right]$, controls the inhomogeneity strength of the plane wave. $\kappa =0$ and $\kappa \ne 0$ correspond to the plane wave being homogeneous and inhomogeneous, respectively. $\kappa $ contributes to an inhomogeneity angle $\vartheta =\arccos \left[ {\operatorname{Im}\left( \tau  \right)}/{\sqrt{{{\operatorname{Im}}^{2}}\left( \tau  \right)+{{\kappa }^{2}}}}\; \right]$ in the range $\left[ -{\pi }/{2}\;,+{\pi }/{2}\; \right]$ between the directions of the propagation and the attenuation. 

In terms of equation (\ref{eq31}), we have
\begin{equation}\tag{32}\label{eq32}
\boldsymbol{s}\cdot \boldsymbol{s}={{s}^{2}}=\left( \tau \boldsymbol{\hat{n}}+\text{i}\kappa \boldsymbol{\hat{m}} \right)\cdot \left( \tau \boldsymbol{\hat{n}}+\text{i}\kappa \boldsymbol{\hat{m}} \right)={{\tau }^{2}}-{{\kappa }^{2}},
\end{equation}
where $\boldsymbol{\hat{l}}\cdot \boldsymbol{\hat{l}}=1$ has been used. We emphasize that the parameter $\kappa $ needs to be given for a certain material, and the complex slowness $s$ is determined by the properties of the material and is solved from the Christoffel equation or the multi-degree algebraic equation. The determination of parameter $\kappa $ means that the complex quantity $\tau $ can be obtained, which is related to the phase velocity ${{V}^{p}}$ by the following relation:
\begin{equation}\tag{33}\label{eq33}
{{V}^{p}}=\frac{1}{\operatorname{Re}\left( \tau  \right)},
\end{equation}
where the symbol $\operatorname{Re}\left( \cdot  \right)$ denotes the real part of a quantity. 

To solve the spatial partial differential in a Cartesian coordinate system, the complex unit vector $\boldsymbol{\hat{l}}$ is expressed in terms of the Cartesian unit vector ${{\boldsymbol{\hat{e}}}_{i}}$ ($i=1,2,3$) in the 3-D case as
\begin{equation}\tag{34}\label{eq34}
\boldsymbol{\hat{l}}={{l}_{1}}{{\boldsymbol{\hat{e}}}_{1}}+{{l}_{2}}{{\boldsymbol{\hat{e}}}_{2}}+{{l}_{3}}{{\boldsymbol{\hat{e}}}_{3}},                   \end{equation}
where
\begin{equation}\tag{35}\label{eq35}
{{l}_{1}}=\frac{{{s}_{1}}}{s}, {{l}_{2}}=\frac{{{s}_{2}}}{s}, \text{and} {{l}_{3}}=\frac{{{s}_{3}}}{s}                  \end{equation}
are defined as the direction cosine in the direction of the complex slowness vector. In the case of inhomogeneous wave propagation, ${{l}_{1}}$, ${{l}_{2}}$, and ${{l}_{3}}$ are complex quantities, but for homogeneous wave propagation, they are real and correspond to the direction cosines of the propagation direction \citep{carcione1992anisotropic}. In this way, equation (\ref{eq30}) can be recast as
\begin{equation}\tag{36}\label{eq36}
\boldsymbol{V}={{\boldsymbol{V}}^{0}}\exp \left[ \text{i}\omega \left( s\left( {{l}_{1}}x+{{l}_{2}}y+{{l}_{3}}z \right)-t \right) \right].
\end{equation}

With equations (\ref{eq36}), the space differential operator matrix (\ref{eq25}) becomes
\begin{equation}\tag{37}\label{eq37}
\boldsymbol{D}=\text{i}\omega \boldsymbol{S}=\text{i}\omega s\boldsymbol{L},  \end{equation}
where the complex matrices $\boldsymbol{S}$ and $\boldsymbol{L}$ can be written as
\begin{equation}\tag{38.1}\label{eq38.1}
\boldsymbol{S}=\left[ \begin{matrix}
   {{s}_{1}} & 0 & 0 & 0 & {{s}_{3}} & {{s}_{2}} & 0 & 0 & 0  \\
   0 & {{s}_{2}} & 0 & {{s}_{3}} & 0 & {{s}_{1}} & 0 & 0 & 0  \\
   0 & 0 & {{s}_{3}} & {{s}_{2}} & {{s}_{1}} & 0 & 0 & 0 & 0  \\
   0 & 0 & 0 & 0 & 0 & 0 & {{s}_{1}} & 0 & 0  \\
   0 & 0 & 0 & 0 & 0 & 0 & 0 & {{s}_{2}} & 0  \\
   0 & 0 & 0 & 0 & 0 & 0 & 0 & 0 & {{s}_{3}}  \\
\end{matrix} \right],
\end{equation}
and
\begin{equation}\tag{38.2}\label{eq38.2}
\boldsymbol{L}=\left[ \begin{matrix}
   {{l}_{1}} & 0 & 0 & 0 & {{l}_{3}} & {{l}_{2}} & 0 & 0 & 0  \\
   0 & {{l}_{2}} & 0 & {{l}_{3}} & 0 & {{l}_{1}} & 0 & 0 & 0  \\
   0 & 0 & {{l}_{3}} & {{l}_{2}} & {{l}_{1}} & 0 & 0 & 0 & 0  \\
   0 & 0 & 0 & 0 & 0 & 0 & {{l}_{1}} & 0 & 0  \\
   0 & 0 & 0 & 0 & 0 & 0 & 0 & {{l}_{2}} & 0  \\
   0 & 0 & 0 & 0 & 0 & 0 & 0 & 0 & {{l}_{3}}  \\
\end{matrix} \right].
\end{equation}

Substituting equation (\ref{eq37}) into equations (\ref{eq27}) and (\ref{eq29}) yields the Christoffel equation
\begin{equation}\tag{39}\label{eq39}
({{\boldsymbol{G}}^{-1}}\cdot \boldsymbol{\Gamma}-{{s}^{-2}}{{\boldsymbol{I}}_{6}})\cdot \boldsymbol{V}=0,  \end{equation}
where 
\begin{equation}\tag{40}\label{eq40}
\boldsymbol{\Gamma}=\boldsymbol{L}\cdot \boldsymbol{R}\cdot {{\boldsymbol{L}}^{\operatorname{T}}}           \end{equation}
is the Christoffel matrix, and ${{\boldsymbol{I}}_{6}}$ represents the sixth-rank identity matrix. The condition that the extended particle velocity vector $\boldsymbol{V}$ has a non-zero solution reads
\begin{equation}\tag{41}\label{eq41}
\det ({{\boldsymbol{G}}^{-1}}\cdot \boldsymbol{\Gamma}-{{s}^{-2}}{{\boldsymbol{I}}_{6}})=0,                     \end{equation}
where the symbol ‘det’ represents the determinant. 

In the analysis of homogeneous plane waves, since $\boldsymbol{L}$ is a real matrix, which is equivalent to a matrix composed of the components of the unit vector $\boldsymbol{\hat{n}}=\left( {{n}_{1}},{{n}_{2}},{{n}_{3}} \right)$ of propagation direction, the matrix $\boldsymbol{\Gamma}$ does not depend on the complex slowness. In this case, equations (\ref{eq39}) and (\ref{eq41}) form an eigensystem and are usually used to determine six eigenvalues and the corresponding eigenvectors. Four of them are responsible for obtaining the complex slowness and the extended particle velocity vector of the qP1, qS1, qS2, and qP2 waves, and the remaining eigenvalue is zero. Nevertheless, for inhomogeneous plane waves, the Christoffel matrix $\boldsymbol{\Gamma}$ is also related to the unknown complex slowness (see equations (\ref{eq35}) and (\ref{eq38.2})). Therefore, it is no longer feasible to solve the eigensystem. 

Under these circumstances, to obtain the complex slowness of the qP1, qS1, qS2, and qP2 waves, an alternative way is to calculate the equation (\ref{eq41}) for given vectors $\boldsymbol{\hat{n}}$, $\boldsymbol{\hat{m}}$, and an inhomogeneity parameter $\kappa $, which leads to deduce an eighth-degree algebraic equation on the complex quantity $\tau $. Standard approaches can be applied in solving the algebraic equation. Once the solution is found, the complex slowness (equation (\ref{eq32})) and the phase velocity (equation (\ref{eq33})) can be easily determined. Substituting the calculated complex slowness into the Christoffel equation (\ref{eq39}), we can solve the corresponding eigenvector $\boldsymbol{V}$ for different wave modes. 

\subsection{Explicit energy velocities and dissipation factors for inhomogeneous waves in attenuative anisotropic Media}\label{sec:theorysub4}
Strictly speaking, in seismology, the spatial position of the wave front at a propagation time is expressed by the energy velocity, and its calculation is more complicated. In poroelastic media, the group velocity and the energy velocity are equivalent. Since the calculation of the group velocity is relatively easy, it is used to represent the position of the wave front. However, in attenuative porous media, the group velocity becomes meaningless due to the dispersion effect caused by poro-viscoelasticity \citep{cheng2021wave}. In this case, the energy velocity needs to be calculated to represent the locus of the wave front accurately. Furthermore, poro-viscoelasticity and WIFF lead to energy loss, which causes the conventional dissipation factor (also called the inverse quality factor) expressions derived from a single-phase medium to be invalid for quantifying attenuation \citep{liu2020seismic, cheng2021wave}. Meanwhile, for inhomogeneous waves propagation, the dissipation factors strongly depend on the degree of wave inhomogeneity, and the dissipation factor expressions derived from the assumption of homogeneous wave propagation are inaccurate. Therefore, it is necessary to deduce an explicit dissipation factor equation, starting from the inhomogeneous waves, to measure the attenuation value in attenuative anisotropic porous media. 

The derivation of the energy velocity and the quality factor follows the energy balance equation, which describes the dynamic process of wave propagation. Actually, \cite{carcione2014wave} presented the complex form of the energy balance equation of poro-viscoelastic anisotropic media. However, due to consideration of the attenuation anisotropy in the porous media, we turn the variation of fluid content into a vector in equation (\ref{eq11}). This causes the equation presented by \cite{carcione2014wave} cannot be directly applied in deriving the expressions of energy velocity and dissipation factor. Hence, we deduce the energy balance equation in attenuative anisotropic porous media, which can be expressed as
\begin{equation}\tag{42}\label{eq42}
\nabla \cdot \boldsymbol{p}=2\text{i}\omega ({{E}_{s}}+{{E}_{v}})-\omega \left( {{E}_{dv}}-{{E}_{ds}} \right).
\end{equation}
The detailed derivation workflow is presented in Appendix \ref{appendA}. In this equation, $\boldsymbol{p}$ denotes the complex Umov–Poynting vector, ${{E}_{s}}$and ${{E}_{v}}$ are the average strain and kinetic energy densities, respectively, ${{E}_{ds}}$ and ${{E}_{dv}}$ are the average dissipated strain and kinetic energy densities, respectively. The expressions of these quantities are given in Appendix \ref{appendA}. Note that equation (\ref{eq42}) is similar to the equation proposed by \cite{carcione2014wave}, the difference lies in the calculation of the average strain and dissipative strain energy densities. When the medium is lossless, the wave energy is not attenuated. If the energy no longer flows into or flows out the closed surface (the source is absent), then $\nabla \cdot \boldsymbol{p}=0$. In this case, in terms of the energy balance equation (\ref{eq42}), we can find that the average strain energy density ${{E}_{s}}$ is equal to the average kinetic energy density ${{E}_{v}}$. 

The energy velocity is defined as the ratio of the average power flow density to the average stored energy density. The real part of the complex Umov–Poynting vector gives the average power flow density. The average stored energy density is identical to the sum of the average strain and kinetic energy densities. Thus, the energy velocity is
\begin{equation}\tag{43}\label{eq43}
{{\boldsymbol{V}}^{e}}=\frac{\operatorname{Re}\left[ \boldsymbol{p} \right]}{{{E}_{v}}+{{E}_{s}}}.
\end{equation}
After the derivation (see Appendix \ref{appendB}), the energy velocity is expressed as
\begin{equation}\tag{44}\label{eq44}
{{\boldsymbol{V}}^{e}}=\frac{\operatorname{Re}\left[ {{\overline{\boldsymbol{V}}}^{\operatorname{T}}}\left( {{{\boldsymbol{\hat{e}}}}_{i}}{{\boldsymbol{\Theta }}_{i}} \right)\cdot \boldsymbol{R}\cdot {{\boldsymbol{S}}^{\text{T}}}\cdot \boldsymbol{V} \right]}{\operatorname{Re}\left[ {{\overline{\boldsymbol{V}}}^{\operatorname{T}}}\cdot \operatorname{Re}\left( \boldsymbol{S} \right)\cdot \boldsymbol{R}\cdot {{\boldsymbol{S}}^{\text{T}}}\cdot \boldsymbol{V} \right]},
\end{equation}
where the matrices ${{\boldsymbol{\Theta }}_{i}}$, $i=1,2,3$ are given in Appendix \ref{appendB}. A definite relationship exists between energy and phase velocities, that is, the phase velocity is the projection of the energy velocity on the propagation direction, which has been proved in single-phase elastic and viscoelastic media \citep{carcione1993energy}, poroelastic media, and the poro-viscoelastic Biot \citep{carcione2014wave} and Biot-squirt (BISQ) media \citep{cheng2021wave}. It still holds in the proposed attenuative anisotropic porous medium, which is verified in Appendix \ref{appendC}.

The dissipation factor is used to represent the anelastic behavior of seismic waves and to characterize the energy transfer in attenuation media. It has two formal definitions in the publications. These definitions are: 
\begin{enumerate}
    \item The dissipation factor (denoted as $Q_{1}^{-1}$) is the ratio of the average dissipated energy density ${{E}_{d}}$ to twice the average strain energy density \citep{carcione2014wave}.
    \item The dissipation factor (denoted as $Q_{2}^{-1}$) is equal to the average dissipated energy density divided by the average stored energy density ${{E}_{a}}$ \citep{buchen1971plane}.
\end{enumerate}
The average dissipated (equation (\ref{eqA12})) and stored energy densities (equation (\ref{eqA11})) are presented in Appendix \ref{appendA}. 

Following these definitions,
$Q_{1}^{-1}$ and $Q_{2}^{-1}$ can be written as
\begin{equation}\tag{45}\label{eq45}
Q_{1}^{-1}=\frac{{{E}_{d}}}{2{{E}_{s}}},       \end{equation}
\begin{equation}\tag{46}\label{eq46}
Q_{2}^{-1}=\frac{{{E}_{d}}}{{{E}_{a}}}.        \end{equation}
By the derivation in Appendix \ref{appendB}, the dissipation factors $Q_{1}^{-1}$ and $Q_{2}^{-1}$ are obtained as
\begin{equation}\tag{47}\label{eq47}
Q_{1}^{-1}=\frac{{{E}_{d}}}{2{{E}_{s}}}=\frac{2\operatorname{Re}\left[ {{\overline{\boldsymbol{V}}}^{\operatorname{T}}}\cdot \operatorname{Im}\left( \boldsymbol{S} \right)\cdot \boldsymbol{R}\cdot {{\boldsymbol{S}}^{\text{T}}}\cdot \boldsymbol{V} \right]}{\operatorname{Re}\left[ {{\overline{\boldsymbol{V}}}^{\operatorname{T}}}\cdot \overline{\boldsymbol{S}}\cdot \boldsymbol{R}\cdot {{\boldsymbol{S}}^{\text{T}}}\cdot \boldsymbol{V} \right]},
\end{equation}
\begin{equation}\tag{48}\label{eq48}
Q_{2}^{-1}=\frac{{{E}_{d}}}{{{E}_{a}}}=\frac{2\operatorname{Re}\left[ {{\overline{\boldsymbol{V}}}^{\operatorname{T}}}\cdot \operatorname{Im}\left( \boldsymbol{S} \right)\cdot \boldsymbol{R}\cdot {{\boldsymbol{S}}^{\text{T}}}\cdot \boldsymbol{V} \right]}{\operatorname{Re}\left[ {{\overline{\boldsymbol{V}}}^{\operatorname{T}}}\cdot \operatorname{Re}\left( \boldsymbol{S} \right)\cdot \boldsymbol{R}\cdot {{\boldsymbol{S}}^{\text{T}}}\cdot \boldsymbol{V} \right]}.
\end{equation}
As can be appreciated in equations (\ref{eq47}) and (\ref{eq48}), the difference between the two dissipation factors comes from the denominator. Using the properties of complex numbers, equation (\ref{eq47}) becomes
\begin{equation}\tag{49}\label{eq49}
Q_{1}^{-1}=\frac{2\operatorname{Re}\left[ {{\overline{\boldsymbol{V}}}^{\operatorname{T}}}\cdot \operatorname{Im}\left( \boldsymbol{S} \right)\cdot \boldsymbol{R}\cdot {{\boldsymbol{S}}^{\text{T}}}\cdot \boldsymbol{V} \right]}{\operatorname{Re}\left[ {{\overline{\boldsymbol{V}}}^{\operatorname{T}}}\cdot \operatorname{Re}\left( \boldsymbol{S} \right)\cdot \boldsymbol{R}\cdot {{\boldsymbol{S}}^{\text{T}}}\cdot \boldsymbol{V} \right]+\operatorname{Im}\left[ {{\overline{\boldsymbol{V}}}^{\operatorname{T}}}\cdot \operatorname{Im}\left( \boldsymbol{S} \right)\cdot \boldsymbol{R}\cdot {{\boldsymbol{S}}^{\text{T}}}\cdot \boldsymbol{V} \right]}.
\end{equation}
From this equation, we can find that the dissipation factor $Q_{1}^{-1}$, compared with $Q_{2}^{-1}$, has one more term
$\operatorname{Im}\left[ {{\overline{\boldsymbol{V}}}^{\operatorname{T}}}\cdot \operatorname{Im}\left( \boldsymbol{S} \right)\cdot \boldsymbol{R}\cdot {{\boldsymbol{S}}^{\text{T}}}\cdot \boldsymbol{V} \right]$ in the denominator. According to the definitions of ${{E}_{s}}$ and ${{E}_{a}}$, this variation depends on the difference between ${{E}_{s}}$ and ${{E}_{v}}$, that is,
\begin{equation}\tag{50}\label{eq50}
\operatorname{Im}\left[ {{\overline{\boldsymbol{V}}}^{\operatorname{T}}}\cdot \operatorname{Im}\left( \boldsymbol{S} \right)\cdot \boldsymbol{R}\cdot {{\boldsymbol{S}}^{\text{T}}}\cdot \boldsymbol{V} \right]={{E}_{s}}-{{E}_{v}}.      \end{equation}

In this section, the explicit expressions for the energy velocity and the dissipation factor of general inhomogeneous waves in attenuative anisotropic porous media are derived from energy considerations. These quantities can be represented in terms of the complex slowness matrix $\boldsymbol{S}$, the extended complex stiffness matrix $\boldsymbol{R}$, and the eigenvectors $\boldsymbol{V}$ of the Christoffel equation, which are relatively clear in structure and easy to be programmed. Using the presented expressions, the wave behavior characteristics in the proposed model can be accurately analyzed.

\subsection{Reduced form of measurable quantities for homogeneous waves}\label{sec:theorysub5}
We have successfully invoked inhomogeneous plane wave theory to derive the expressions of measurable quantities in the attenuative anisotropic porous media. In this section, we present the reduced form of the expressions for the energy velocities and the dissipation factors in the special case of the homogeneous wave, and compare it with the existing expressions derived by using wave vector $\boldsymbol{k}$. 

As stated before, when $\kappa =0$, the wave propagation is homogeneous, then the matrix $\boldsymbol{L}$ is a real matrix. On this condition, using equations (\ref{eq37}), (\ref{eq39}), and (\ref{eq40}), the energy velocity becomes
\begin{equation}\tag{51}\label{eq51}
{{\boldsymbol{V}}^{e}}=\frac{\operatorname{Re}\left[ s{{\boldsymbol{V}}^{\operatorname{T}}}\cdot \boldsymbol{L}\cdot {{\boldsymbol{R}}^{T}}\cdot \left( {{{\boldsymbol{\hat{e}}}}_{i}}\boldsymbol{\Theta }_{i}^{\operatorname{T}} \right)\cdot \overline{\boldsymbol{V}} \right]}{\operatorname{Re}\left[ \operatorname{Re}\left( s \right){{s}^{-1}}{{\boldsymbol{V}}^{\operatorname{T}}}\cdot \boldsymbol{G}\cdot \overline{\boldsymbol{V}} \right]}.
\end{equation}
Here, the properties
${{\overline{\boldsymbol{V}}}^{\operatorname{T}}}\left( {{{\boldsymbol{\hat{e}}}}_{i}}{{\boldsymbol{\Theta }}_{i}} \right)\cdot \boldsymbol{R}\cdot {{\boldsymbol{L}}^{\text{T}}}\cdot \boldsymbol{V}={{\boldsymbol{V}}^{\operatorname{T}}}\cdot \boldsymbol{L}\cdot {{\boldsymbol{R}}^{T}}\cdot \left( {{{\boldsymbol{\hat{e}}}}_{i}}\boldsymbol{\Theta }_{i}^{\operatorname{T}} \right)\cdot \overline{\boldsymbol{V}}$ and ${{\overline{\boldsymbol{V}}}^{\operatorname{T}}}\cdot \boldsymbol{G}\cdot \boldsymbol{V}={{\boldsymbol{V}}^{\operatorname{T}}}\cdot \boldsymbol{G}\cdot \overline{\boldsymbol{V}}$ have been used. Equation (\ref{eq51}) is identical to the energy velocity expressions derived in the Biot model with the RF mechanism \citep{carcione1996wave} and the BISQ model with the RS mechanism \citep{cheng2021wave} under the case of the homogeneous wave as follows
\begin{equation}\tag{52}\label{eq52}
{{\boldsymbol{V}}^{e}}=\frac{2\operatorname{Re}\left[ {{V}^{-1}}{{\boldsymbol{V}}^{\operatorname{T}}}\cdot \boldsymbol{L}\cdot {{\boldsymbol{R}}^{T}}\cdot \left( {{{\boldsymbol{\hat{e}}}}_{i}}\boldsymbol{\Theta }_{i}^{\operatorname{T}} \right)\cdot \overline{\boldsymbol{V}} \right]}{\operatorname{Re}\left[ (1+{{\left| V \right|}^{-2}}{{V}^{2}}){{\boldsymbol{V}}^{\operatorname{T}}}\cdot \boldsymbol{G}\cdot \overline{\boldsymbol{V}} \right]},
\end{equation}
where $V={\omega }/{k}\;={{s}^{-1}}$ is the complex velocity. Note that, the matrices $\boldsymbol{L}\cdot {{\boldsymbol{R}}^{\text{T}}}$ and $\boldsymbol{G}$ in equation (\ref{eq52}) are represented by the symbols ${{\boldsymbol{C}}^{\text{T}}}$ and $\boldsymbol{\Gamma}$ respectively in the expression given by \cite{carcione1996wave}.

On the other hand, from equations (\ref{eq37}), (\ref{eq39}), and (\ref{eq40}), equations (\ref{eq49}) and (\ref{eq48}) for the homogeneous wave reduce to
\begin{equation}\tag{53}\label{eq53}
Q_{1\text{h}}^{-1}=\frac{2\operatorname{Im}\left( s \right)}{\operatorname{Re}\left( s \right)+\operatorname{Im}\left( s \right)\frac{\operatorname{Im}\left[ s{{\overline{\boldsymbol{V}}}^{\operatorname{T}}}\cdot \boldsymbol{\Gamma}\cdot \boldsymbol{V} \right]}{\operatorname{Re}\left[ s{{\overline{\boldsymbol{V}}}^{\operatorname{T}}}\cdot \boldsymbol{\Gamma}\cdot \boldsymbol{V} \right]}},
\end{equation}
\begin{equation}\tag{54}\label{eq54}
Q_{2\text{h}}^{-1}=\frac{2\operatorname{Im}\left( s \right)}{\operatorname{Re}\left( s \right)}.
\end{equation}
Equation (\ref{eq53}) is consistent with the dissipation factor expression of the homogeneous wave in the Biot model with the RF mechanism and the BISQ model with the RS mechanism as follows
\begin{equation}\tag{55}\label{eq55}
{{Q}^{-1}}=\frac{2\operatorname{Im}\left[ V \right]\operatorname{Re}\left[ V{{\boldsymbol{V}}^{\operatorname{T}}}\cdot \boldsymbol{G}\cdot \overline{\boldsymbol{V}} \right]}{\operatorname{Re}\left[ {{V}^{2}}{{\boldsymbol{V}}^{\operatorname{T}}}\cdot \boldsymbol{G}\cdot \overline{\boldsymbol{V}} \right]},
\end{equation}
which is presented by \cite{carcione2014wave} and \cite{cheng2021wave} using the definition of $Q_{1}^{-1}$. Interestingly, equation (\ref{eq54}) is the same as the expression ${{Q}^{-1}}={2\operatorname{Im}\left( k \right)}/{\operatorname{Re}\left( k \right)}\;$ \citep{parra1997transversely, yang2002poroelastic, ba2017rock} obtained by a high-$Q$ approximation of the dissipation factor expression ${{Q}^{-1}}={\operatorname{Im}\left( {{V}^{2}} \right)}/{\operatorname{Re}\left( {{V}^{2}} \right)}\;$ in the single-phase media due to $k=\omega s$. This indirectly implies that the dissipation factor expression derived from the assumption of the homogeneous wave is not desirable to quantify the attenuation value in the attenuative porous media.

It is demonstrated that the energy velocities and dissipation factors expressions of the inhomogeneous waves can be perfectly degenerated to that of the homogeneous waves as a special case. This implies the presented expressions are more general. It needs to be emphasized that although the expressions are given for different media, they are found to have a unified form. The essential difference between them comes from the material properties associated with each term in the expressions. 

%% file: Sections/Numerical_Examples.tex
\section{\textbf{Numerical examples}}
The following examples are intended to analyze the propagation of inhomogeneous plane waves in an attenuating anisotropic medium. We consider a porous medium with VTI attenuation used to describe shales or other layered rocks. First, we validate the influences of the attenuation anisotropy, the RS mechanism, and the RF mechanism on the wave behavior. Then, how attenuation anisotropy and wave inhomogeneity affect the dispersion and attenuation of seismic waves is studied. Finally, the differences between the two dissipation factors under different definitions are presented. In this section, we assume that all porous materials are saturated by water, the common properties are shown in Table \ref{tab1}, and the reference frequency ${{\omega }_{0}}$ is 30 Hz.

We emphasize that, under consideration of attenuative VTI porous media, the calculation assumes propagation of inhomogeneous plane waves in the plane of symmetry (i.e., the plane $({{x}_{1}},{{x}_{3}})$) without loss of generality. This means that the components of the unit vector $\boldsymbol{\hat{n}}$ can be represented as
\begin{equation}\tag{56}\label{eq56}
{{n}_{1}}=\sin \theta, ~~{{n}_{2}}=0, ~~ \text{and} ~~{{n}_{3}}=\cos \theta,
\end{equation}
where $\theta $ is the propagation angle, which denotes the angle between the wave propagation direction and the positive direction of the $z$-axis. As $\boldsymbol{\hat{n}}\cdot \boldsymbol{\hat{m}}=0$, the components of unit vector $\boldsymbol{\hat{m}}$ become ${{m}_{1}}=\cos \theta $, ${{m}_{2}}=0$, and ${{m}_{3}}=-\sin \theta $.

\begin{table}
  \centering
  \caption{The common properties of solid skeleton and fluid}
  \label{tab1}
  \begin{tabular}{@{} cccccc @{}}
    \toprule
    $K_s$ (GPa) & $K_f$ (GPa) & $\rho_s$ (kg/m$^3$) & $\rho_f$ (kg/m$^3$) & $\varphi$ & $\eta$ (Pa $\cdot$ s) \\
    \midrule
    40          & 2.5         & 2500                & 1040                & 0.2       & 0.001         \\
    \bottomrule
  \end{tabular}
\end{table}

\begin{figure*}
\centering
\includegraphics[width=1\textwidth]{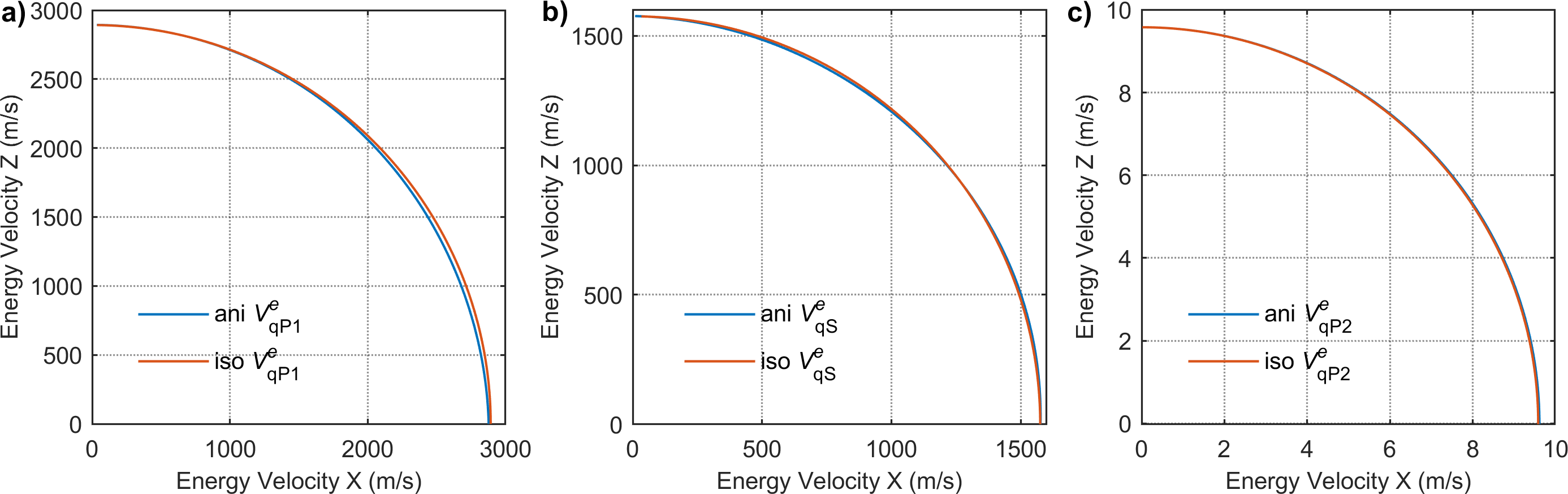}
\caption{Polar representations of the energy velocity curves for inhomogeneous waves in the attenuative VTI porous material and the attenuative isotropic porous material at 50 Hz, where (a), (b), and (c) correspond to the qP1, qS, and qP2 waves, respectively. Here, the inhomogeneity parameter $\kappa $ is $2*{{10}^{-4}}$, and only a quarter of the curves are shown due to symmetry considerations. In (a), (b), and (c), the symbols $V_{\text{qP1}}^{e}$, $V_{\text{qS}}^{e}$, and $V_{\text{qP2}}^{e}$ represent the energy velocity of qP1 wave, qS wave, and qP2 wave, respectively, and the symbols ‘ani’ and ‘iso’ denote the attenuative VTI porous material and the attenuative isotropic porous material, respectively. }
\label{fig1}
\end{figure*}

\begin{figure*}
\centering
\includegraphics[width=1\textwidth]{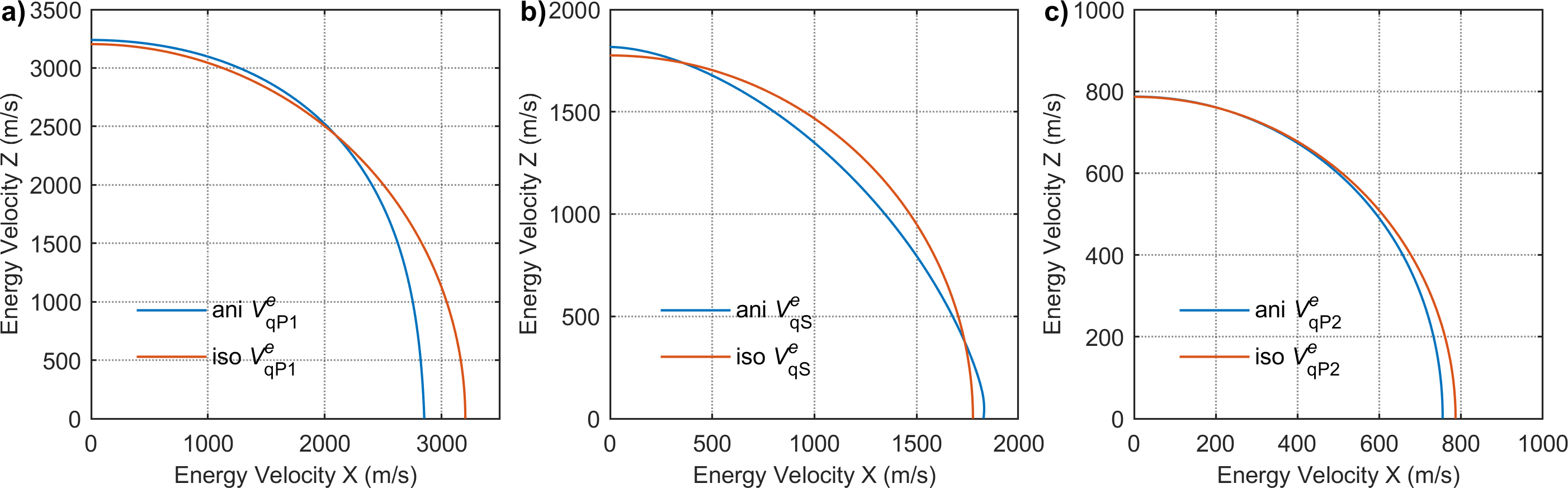}
\caption{Same as Figure \ref{fig1}, but at the frequency of ${{10}^{6}}$ Hz.}
\label{fig2}
\end{figure*}

\begin{figure*}
\centering
\includegraphics[width=1\textwidth]{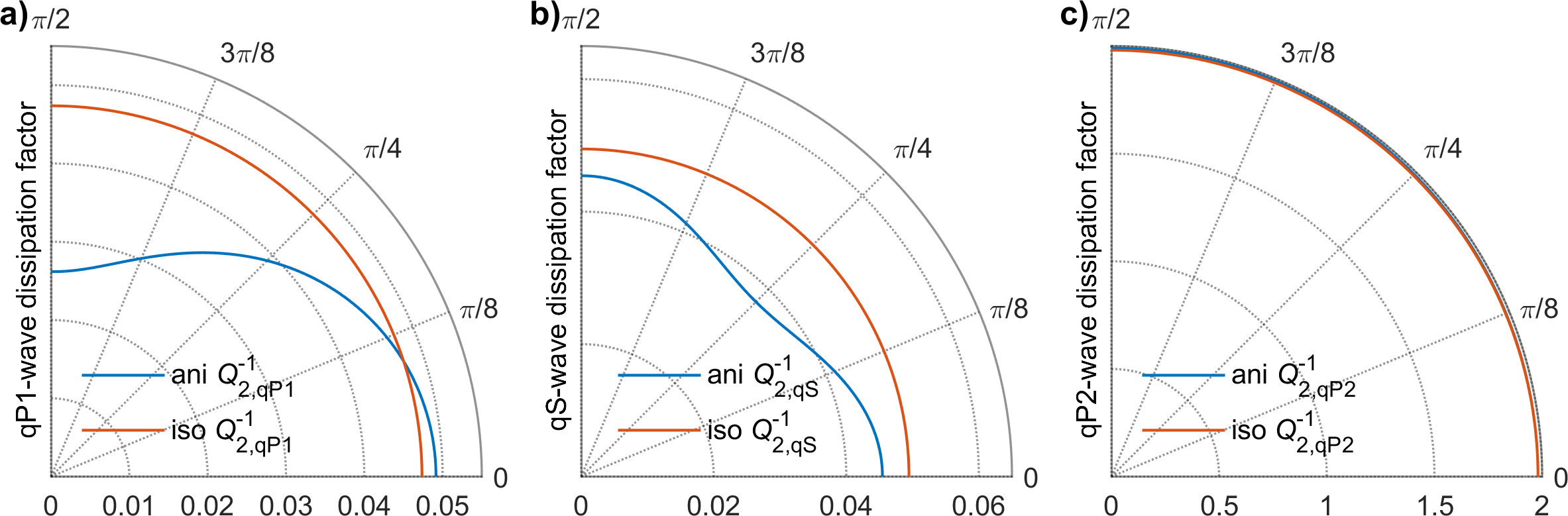}
\caption{Polar representations of the dissipation factor curves for inhomogeneous waves in the attenuative VTI porous material and the attenuative isotropic porous material at 50 Hz, where (a), (b), and (c) correspond to the qP1, qS, and qP2 waves, respectively. Here, the inhomogeneity parameter $\kappa $ is $2*{{10}^{-4}}$, and only a quarter of the curves are shown due to symmetry considerations. In (a), (b), and (c), the symbols $Q_{\text{2,qP1}}^{\text{-1}}$, $Q_{\text{2,qS}}^{\text{-1}}$, and $Q_{\text{2,qP2}}^{\text{-1}}$ represent the energy velocity of qP1 wave, qS wave, and qP2 wave, respectively, and the symbols ‘ani’ and ‘iso’ denote the attenuative VTI porous material and the attenuative isotropic porous material, respectively. }
\label{fig3}
\end{figure*}

\begin{figure*}
\centering
\includegraphics[width=1\textwidth]{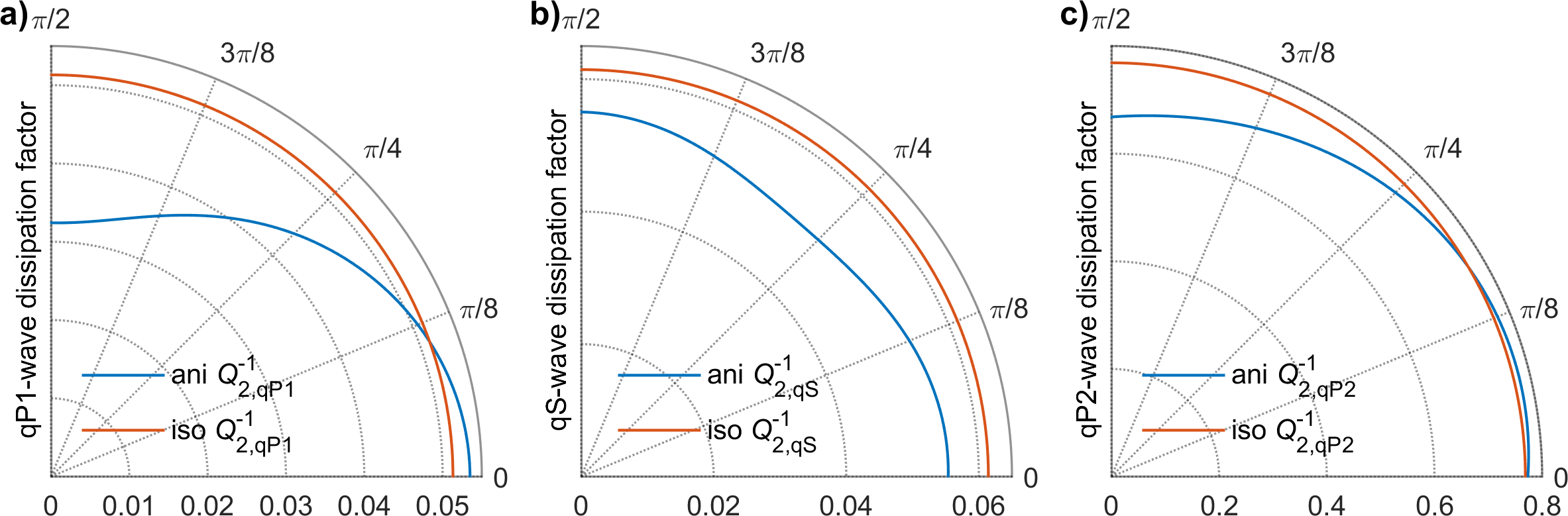}
\caption{Same as Figure \ref{fig3}, but at the frequency of ${{10}^{6}}$ Hz.}
\label{fig4}
\end{figure*}

\begin{figure*}
\centering
\includegraphics[width=1\textwidth]{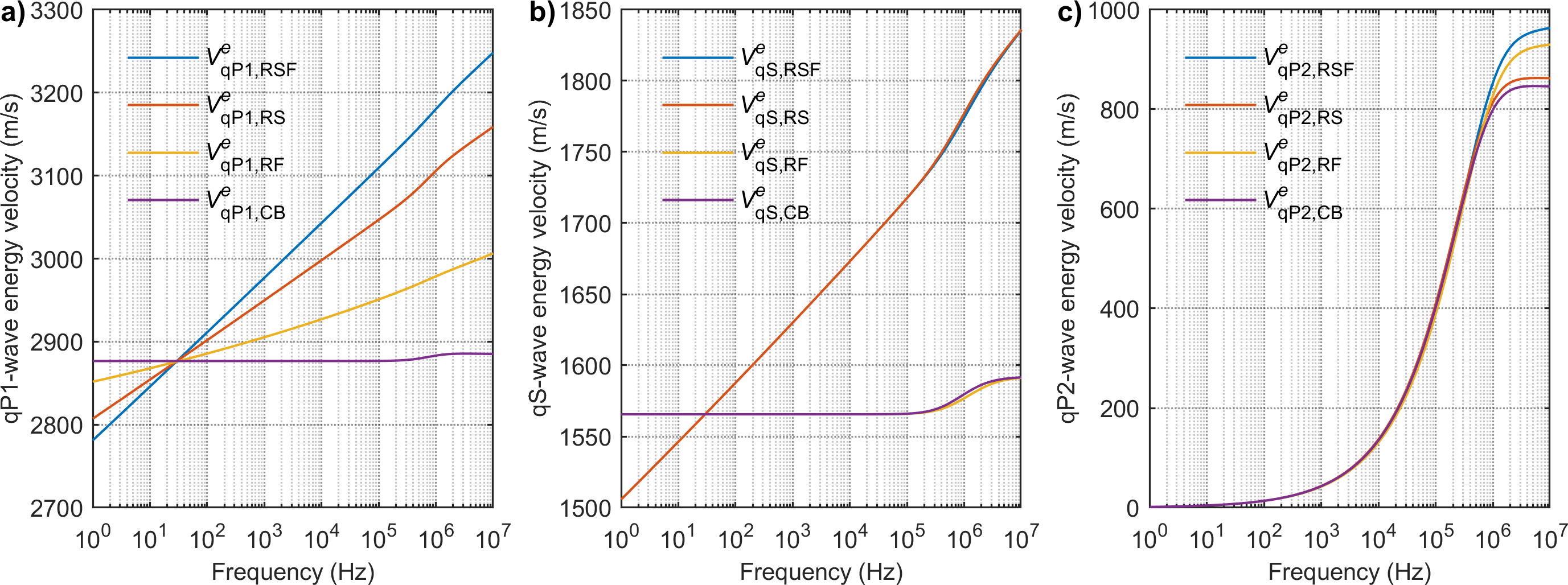}
\caption{Dispersion curves of inhomogeneous waves in the attenuative isotropic porous material (denoted with symbol RSF), relaxed skeleton porous material (denoted with symbol RS), relaxed fluid diffusion porous material (denoted with symbol RF), and classic Biot material (denoted with symbol CB), where (a), (b), and (c) correspond to qP1, qS, and qP2 waves, respectively. Here, the propagation direction is ${\pi }/{\text{4}}\;$, and the inhomogeneity parameter $\kappa $ is $2*{{10}^{-4}}$. In (a), the symbols $V_{\text{qP1,RSF}}^{e}$, $V_{\text{qP1,RS}}^{e}$, $V_{\text{qP1,RF}}^{e}$, and $V_{\text{qP1,CB}}^{e}$ represent the qP1-wave energy velocities in proposed, RS, RF, and CB models, respectively. Panels (b) and (c) are denoted in a manner similar to panel (a).}
\label{fig5}
\end{figure*}

\begin{figure*}
\centering
\includegraphics[width=1\textwidth]{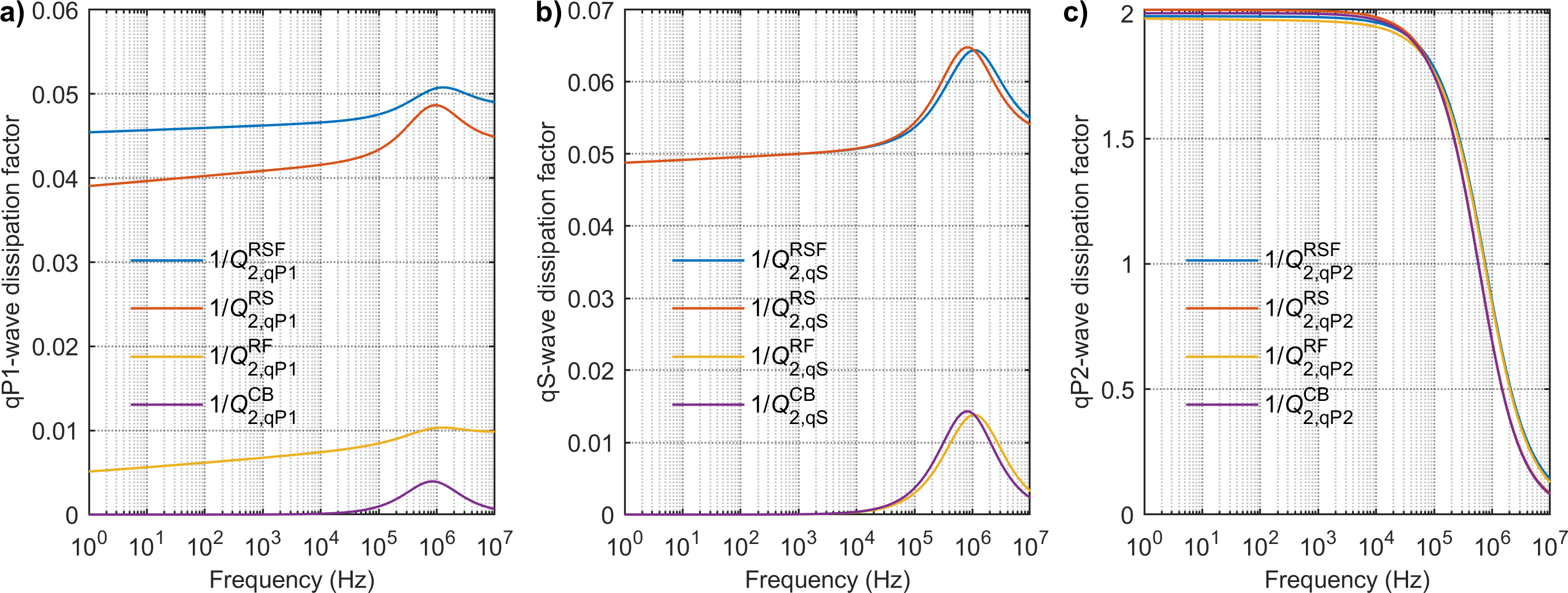}
\caption{Dissipation factors of inhomogeneous waves in the attenuative isotropic porous material (denoted with symbol RSF), relaxed skeleton porous material (denoted with symbol RS), relaxed fluid diffusion porous material (denoted with symbol RF), and classic Biot material (denoted with symbol CB), where (a), (b), and (c) correspond to qP1, qS, and qP2 waves, respectively. Here, the propagation direction is ${\pi }/{\text{4}}\;$, and the inhomogeneity parameter $\kappa $ is $2*{{10}^{-4}}$. In (a), the symbols ${1}/{Q_{\text{2,qP1}}^{\text{RSF}}}\;$, ${1}/{Q_{\text{2,qP1}}^{\text{RS}}}\;$, ${1}/{Q_{\text{2,qP1}}^{\text{RF}}}\;$, and ${1}/{Q_{\text{2,qP1}}^{\text{CB}}}\;$ represent the qP1-wave dissipation factors under the definition of $Q_{2}^{-1}$ in the proposed, RS, RF, and CB models, respectively. Panels (b) and (c) are denoted in a manner similar to panel (a).}
\label{fig6}
\end{figure*}

\subsection{Analysis of the proposed model}
To analyze the attenuation anisotropy of the proposed model, we first calculate the energy velocity curves and the dissipation factor curves in the attenuative VTI porous material and attenuative isotropic porous material, respectively. Here, for the attenuative VTI porous material, the components of the matrices ${{\boldsymbol{Q}}^{s}}$ and ${{\boldsymbol{Q}}^{f}}$ representing the attenuation anisotropy are: $Q_{11}^{s}=50$, $Q_{33}^{s}=25$, $Q_{13}^{s}=40$, $Q_{55}^{s}=25$, $Q_{1}^{f}=40$, and $Q_{3}^{f}=20$. In the porous materials with VTI attenuation, the attenuation of the symmetry axis is greater than that of the horizontal, hence, $Q_{33}^{s}<Q_{11}^{s}$, and it has been assumed that the RS and RF mechanisms have the same attenuation anisotropy (i.e.,${Q_{11}^{s}}/{Q_{33}^{s}}\;={Q_{1}^{f}}/{Q_{3}^{f}}\;$). The attenuative isotropic porous material is obtained from the degradation of attenuative VTI porous material, that is, $Q_{11}^{s}=Q_{33}^{s}=25$, $Q_{55}^{s}=25$, $Q_{1}^{f}=Q_{3}^{f}=20$, and $Q_{13}^{s}=25$ (computed by $Q_{13}^{s}=Q_{33}^{s}{\left( {{C}_{33}}-2{{C}_{55}} \right)}/{\left( {{C}_{33}}-2{{C}_{55}}{Q_{33}^{s}}/{Q_{55}^{s}}\; \right)}\;$). To avoid the influence of velocity anisotropy on the dissipation factor, the drained stiffness matrices of the two materials are the same and both are isotropic, where ${{C}_{11}}={{C}_{33}}=23$$~\text{GPa}$, ${{C}_{55}}=6$$~\text{GPa}$, ${{C}_{12}}={{C}_{13}}={{C}_{33}}-2{{C}_{55}}=11$$~\text{GPa}$. The permeabilities and tortuosity of the two materials are ${{K}_{11}}={{K}_{33}}=100$$~\text{mD}$ and ${{T}_{1}}={{T}_{3}}=3.2$, respectively. In this section, the inhomogeneity parameter $\kappa $ of the plane wave is $2*{{10}^{-4}}$.

In the given two materials, the polar representations of the energy velocity curves for the inhomogeneous waves at 50 Hz and ${{10}^{6}}$ Hz are plotted in Figures \ref{fig1} and \ref{fig2}, respectively, where the blue and orange lines correspond to attenuative VTI porous materials and attenuative isotropic materials, respectively. As displayed in Figures \ref{fig1} and \ref{fig2}, the energy velocity curves of the qP1, qS, and qP2 waves in the attenuative isotropic porous material are circular. For the attenuative VTI porous materials, the change in the energy velocity of each wave caused by the attenuation anisotropy at low frequencies is small but cannot be ignored (Figure \ref{fig1}). Since the energy velocity curves represent the sections of the wavefronts, the observed difference means that the attenuation anisotropy causes the wavefront to propagate to a larger or smaller spatial location. For example, in Figure \ref{fig1}a, the propagation of the qP1-wave wavefront in the attenuative VTI porous material is slower than that in the attenuative isotropic porous material. The effects of attenuation anisotropy on energy velocities are substantial at high frequency (see Figure \ref{fig2}). Even if we assume that the attenuative VTI porous material is velocity isotropic (that is, the drained stiffness components are considered as isotropic), the attenuation anisotropy still significantly affects the shape of the wavefronts of the qP1, qS, and qP2 waves, causing them elliptical.

Figures \ref{fig3} and \ref{fig4} show the dissipation factor curves for the inhomogeneous waves in two materials at 50 Hz and ${{10}^{6}}$ Hz, respectively, where we use the definition of $Q_{2}^{-1}$ to calculate dissipation factors. Similarly, the dissipation factor curves of the qP1, qS, and qP2 waves in the attenuative isotropic porous material are also circular. At both low and high frequencies, the attenuation anisotropy has an obvious impact on the dissipation factors of the qP1 and qS waves, causing them direction-dependent (Figures \ref{fig3}a, \ref{fig3}b, \ref{fig4}a, and \ref{fig4}b). Here is a noteworthy phenomenon, attenuation anisotropy of qS waves is more obvious at low frequencies than at high frequencies. This implies that the variation of frequency enables to change the direction dependence of the dissipation factor in the attenuative VTI porous material. The qP2 wave is a diffusive wave at low frequency, and the attenuation anisotropy hardly affects the dissipation factor (Figure \ref{fig3}c). At high frequency, the qP2 wave is transformed to a propagative wave, hence, the variation in the qP2-wave dissipation factor caused by the attenuation anisotropy can be clearly observed (Figure \ref{fig4}c).

To illustrate the proposed model, which includes both RS and RF mechanisms, can cause more significant attenuation and dispersion in a wide frequency range, we compare it with the velocity dispersion curves and dissipation factor curves of the RS (poro-viscoelastic model considering only the RS mechanism), RF (poro-viscoelastic model considering only the RF mechanism), and the classical Biot (denoted with CB) models in the frequency range $\left[ {{10}^{0}},{{10}^{7}} \right]$. Here, the previous attenuative isotropic porous material is used, but the attenuation matrix components controlling the RS and RF mechanisms become the same, that is, $Q_{11}^{s}=Q_{33}^{s}=Q_{13}^{s}=Q_{55}^{s}=Q_{1}^{f}=Q_{3}^{f}=25$. The RS, RF, and CB models are obtained from its degradation: $Q_{1}^{f}=Q_{3}^{f}=\infty $ (RS model); $Q_{11}^{s}=Q_{33}^{s}=Q_{13}^{s}=Q_{55}^{s}=\infty $ (RF model); $Q_{11}^{s}=Q_{33}^{s}=Q_{13}^{s}=Q_{55}^{s}=\infty $ and $Q_{1}^{f}=Q_{3}^{f}=\infty $(CB model), where $\infty $ is taken as ${{10}^{30}}$. It is emphasized that here the attenuative VTI porous material is not selected to avoid the influence of attenuation anisotropy, i.e., compared with the RF model, the skeleton of the attenuative VTI porous material is not only relaxed but also attenuation anisotropic. In this case, the comparison between the two is no longer be fair.

Figures \ref{fig5}a, \ref{fig5}b, and \ref{fig5}c display the energy velocity curves of the inhomogeneous qP1 ($V_{\text{qP1,RSF}}^{e}$, $V_{\text{qP1,RS}}^{e}$, $V_{\text{qP1,RF}}^{e}$, and $V_{\text{qP1,CB}}^{e}$), qS ($V_{\text{qS,RSF}}^{e}$, $V_{\text{qS,RS}}^{e}$, $V_{\text{qS,RF}}^{e}$, and $V_{\text{qS,CB}}^{e}$), and qP2 ($V_{\text{qP2,RSF}}^{e}$, $V_{\text{qP2,RS}}^{e}$, $V_{\text{qP2,RF}}^{e}$, $V_{\text{qP2,CB}}^{e}$) waves in the proposed model (denoted with subscript RSF), RS model (denoted with subscript RS), RF model (denoted with subscript RF), and CB model (denoted with subscript CB), respectively. For the qP1 wave, the proposed model causes the strongest dispersion, and the dispersion curves $V_{\text{qP1,CB}}^{e}$ have minimal frequency dependence (Figure \ref{fig5}a). The velocity dispersion of the qP1 wave as predicted by the RS model is stronger than that resulting from the RF model. In Figure \ref{fig5}b, we observe that the dispersion curves of $V_{\text{qS,RSF}}^{e}$ and $V_{\text{qS,RS}}^{e}$ are remarkably similar, and the dispersion curves of $V_{\text{qS,RF}}^{e}$ and $V_{\text{qS,CB}}^{e}$ are close to coincidence, which indicates that the RF mechanism scarcely affects the velocity dispersion of the qS wave, as expected. There is a negligible difference in the qP2-wave energy velocity curves of the four models in the frequency bands less than the Biot relaxation frequency around $2.5*{{10}^{5}}$Hz (Figure \ref{fig5}c). This is because when the frequency is less than the Biot relaxation frequency, the fluid flow is dominated by the viscous force. If the frequency is greater than Biot relaxation frequency, it becomes an inertial-force-dominated flow \citep{biot1962generalized, biot1962mechanics}. In this case, the relaxation mechanism will work. Therefore, Figure \ref{fig5}c indicates an evident difference of the qP2-wave energy velocity curves between the four models in the frequency bands greater than the Biot relaxation frequency. The dispersion in the terms $V_{\text{qP2,RSF}}^{e}$ and $V_{\text{qP2,CB}}^{e}$ are the strongest and the weakest, respectively, while the opposite behavior occurs in the RS and RF models compared with Figure \ref{fig5}a.

Figure \ref{fig6} presents the comparison of dissipation factor of the inhomogeneous waves between the proposed model (denoted with superscript RSF), RS model (denoted with superscript RS), RF model (denoted with subscript RF), and CB model (denoted with superscript CB), where a, b and c correspond to the qP1 (${1}/{Q_{\text{2,qP1}}^{\text{RSF}}}\;$, ${1}/{Q_{\text{2,qP1}}^{\text{RS}}}\;$, ${1}/{Q_{\text{2,qP1}}^{\text{RF}}}\;$, and ${1}/{Q_{\text{2,qP1}}^{\text{CB}}}\;$), qS (${1}/{Q_{\text{2,qS}}^{\text{RSF}}}\;$, ${1}/{Q_{\text{2,qS}}^{\text{RS}}}\;$, ${1}/{Q_{\text{2,qS}}^{\text{RF}}}\;$, and ${1}/{Q_{\text{2,qS}}^{\text{CB}}}\;$), and qP2 (${1}/{Q_{\text{2,qP2}}^{\text{RSF}}}\;$, ${1}/{Q_{\text{2,qP2}}^{\text{RS}}}\;$, ${1}/{Q_{\text{2,qP2}}^{\text{RF}}}\;$, and ${1}/{Q_{\text{2,qP2}}^{\text{CB}}}\;$) waves, respectively. When the components of the attenuation matrices describing the RS and RF mechanisms are the same, the RS mechanism attenuates the qP1-wave energy more strongly than the RF mechanism, and the difference between the attenuation values predicted by the two models is enormous. Incorporating both mechanisms, the proposed model yields the strongest attenuation for the qP1 wave. In contrast, the CB model substantially underestimates the attenuation value of the qP1 wave in a wide frequency range (Figure \ref{fig6}a). The qS-wave dissipation factors of ${1}/{Q_{\text{2,qS}}^{\text{RSF}}}\;$ and ${1}/{Q_{\text{2,qS}}^{\text{RS}}}\;$ are slightly different, but they are prominently larger than the virtually identical ${1}/{Q_{\text{2,qS}}^{\text{RF}}}\;$ and ${1}/{Q_{\text{2,qS}}^{\text{CB}}}\;$, which means that the RF mechanism barely contributes to the energy attenuation of the qS wave (Figure \ref{fig6}b), as the same effects on dispersion. Figure \ref{fig6}c shows that the dissipation factor of the qP2 wave has few differences among the four models. The qP2-wave dissipation factors of the four models remain a constant value of 2 in the frequency band less than about ${{10}^{4}}$ Hz. This high dissipation factor implies that the qP2 wave is a diffusive wave. Once the Biot relaxation frequency is exceeded, the dissipation factors gradually decrease to a smaller value, and the qP2 wave becomes a propagative wave. In the high frequencies, the attenuation values measured by the proposed model and the RF model are slightly larger than those predicted by the RS and CB models. Note that the RS mechanism hardly attenuates the qP2-wave energy. Thus, the dissipation factor curves of ${1}/{Q_{\text{2,qP2}}^{\text{RS}}}\;$ and ${1}/{Q_{\text{2,qP2}}^{\text{CB}}}\;$ are almost coincident, and the attenuation values resulting from the proposed model and the RF model are nearly equal. 

From the above examples, we can see that the attenuation anisotropy markedly affects the direction-dependent characteristics of the dissipation factor of seismic waves and changes the shape of the wavefronts more strongly as the frequency increases. The CB model provide a reduced prediction of attenuation value and dispersion. The RF model motivates the development of the CB model, but the improvement is far from enough. The RS model can better explain the frequency dependence of the qP1- and qS-wave velocity, and present a stronger attenuation effect. Compared with the RS, RF, and CB models, the proposed model comprehensively considers the complete RS and RF mechanisms in porous materials. As a result, it enables to describe a stronger dispersion and attenuation of seismic waves in a wide frequency range, especially for qP1 waves concerned in seismic exploration. 

\begin{figure}
\centering
\includegraphics[width=0.66\textwidth]{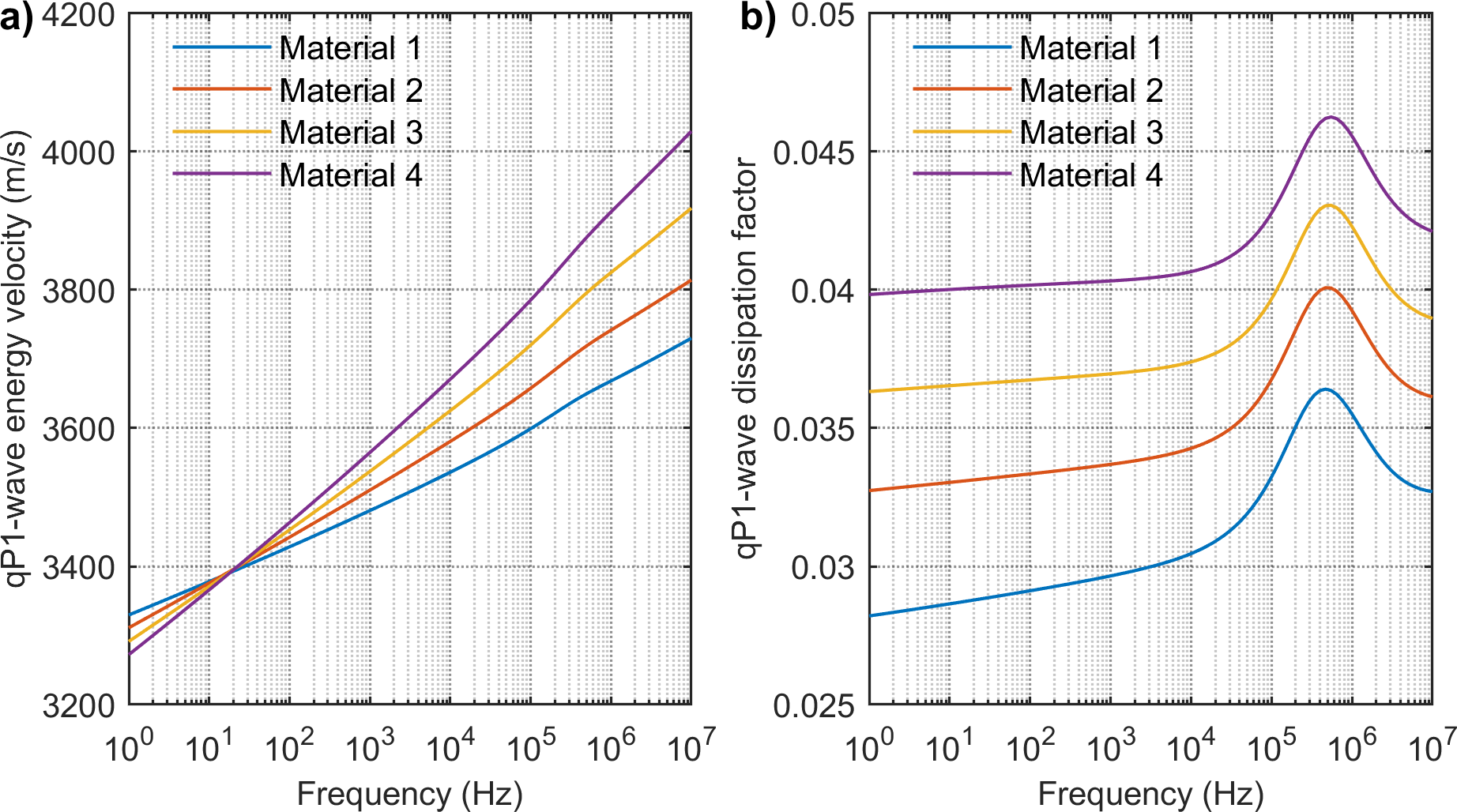}
\caption{Dispersion curve (a) and dissipation factor (b) of inhomogeneous qP1 wave in the attenuative VTI porous material, where the inhomogeneity parameter $\kappa $ is $1*{{10}^{-4}}$, and the propagation angle is ${\pi }/{4}\;$. Here, the properties of materials 1-4 are given in Tables \ref{tab1}, \ref{tab2}, and \ref{tab3}.}
\label{fig7}
\end{figure}

\begin{figure}
\centering
\includegraphics[width=0.66\textwidth]{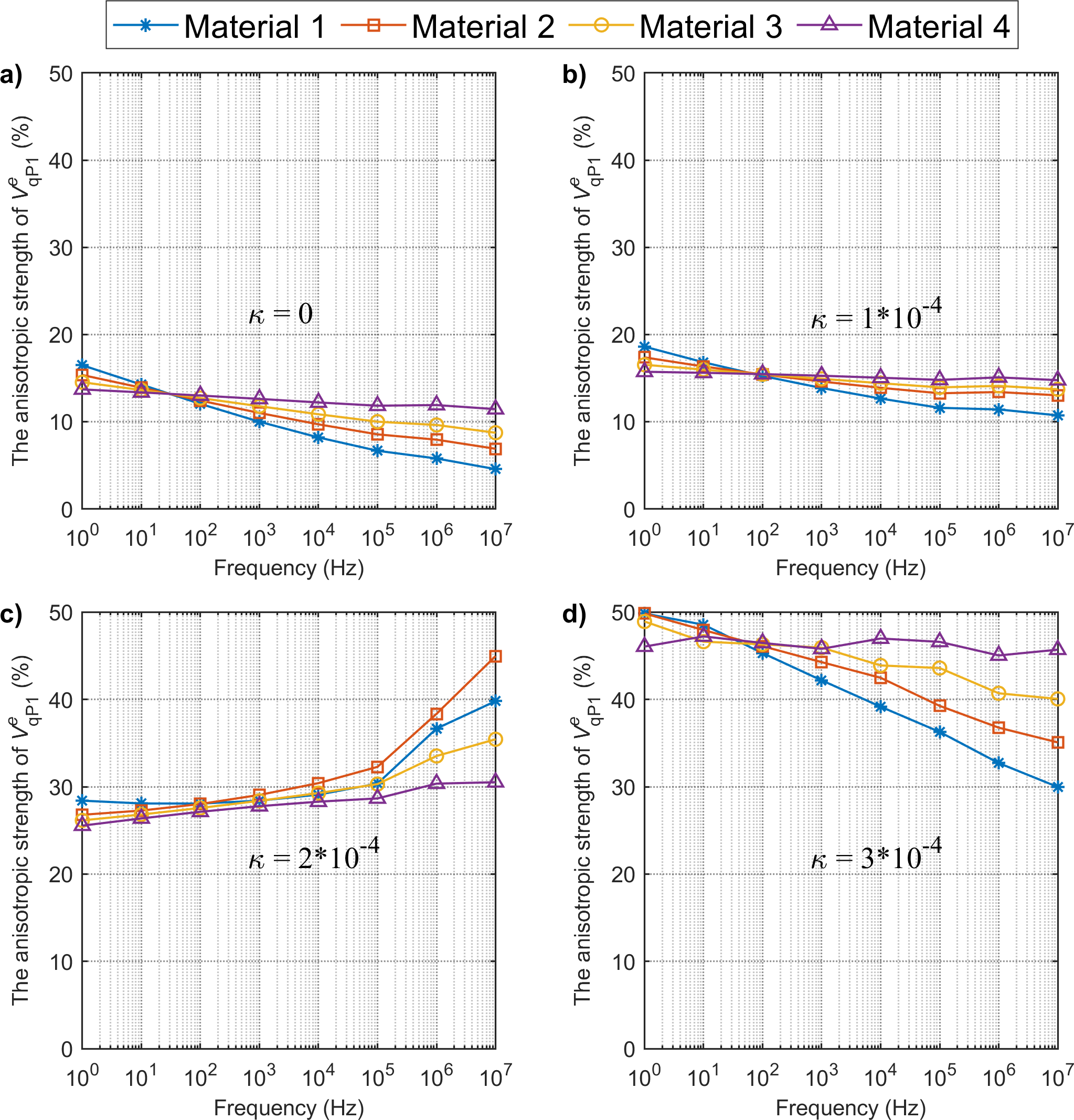}
\caption{The anisotropic factor curves of the qP1-wave energy velocity, where (a), (b), (c), and (d), correspond to inhomogeneity parameter $\kappa =0$, $1*{{10}^{-4}}$, $2*{{10}^{-4}}$, and $3*{{10}^{-4}}$, respectively. The anisotropic factors are calculated at frequencies $1$, ${{10}^{1}}$, ${{10}^{2}}$, ${{10}^{3}}$, ${{10}^{4}}$, ${{10}^{5}}$, ${{10}^{6}}$, and ${{10}^{7}}$, respectively.}
\label{fig8}
\end{figure}

\begin{figure}
\centering
\includegraphics[width=0.66\textwidth]{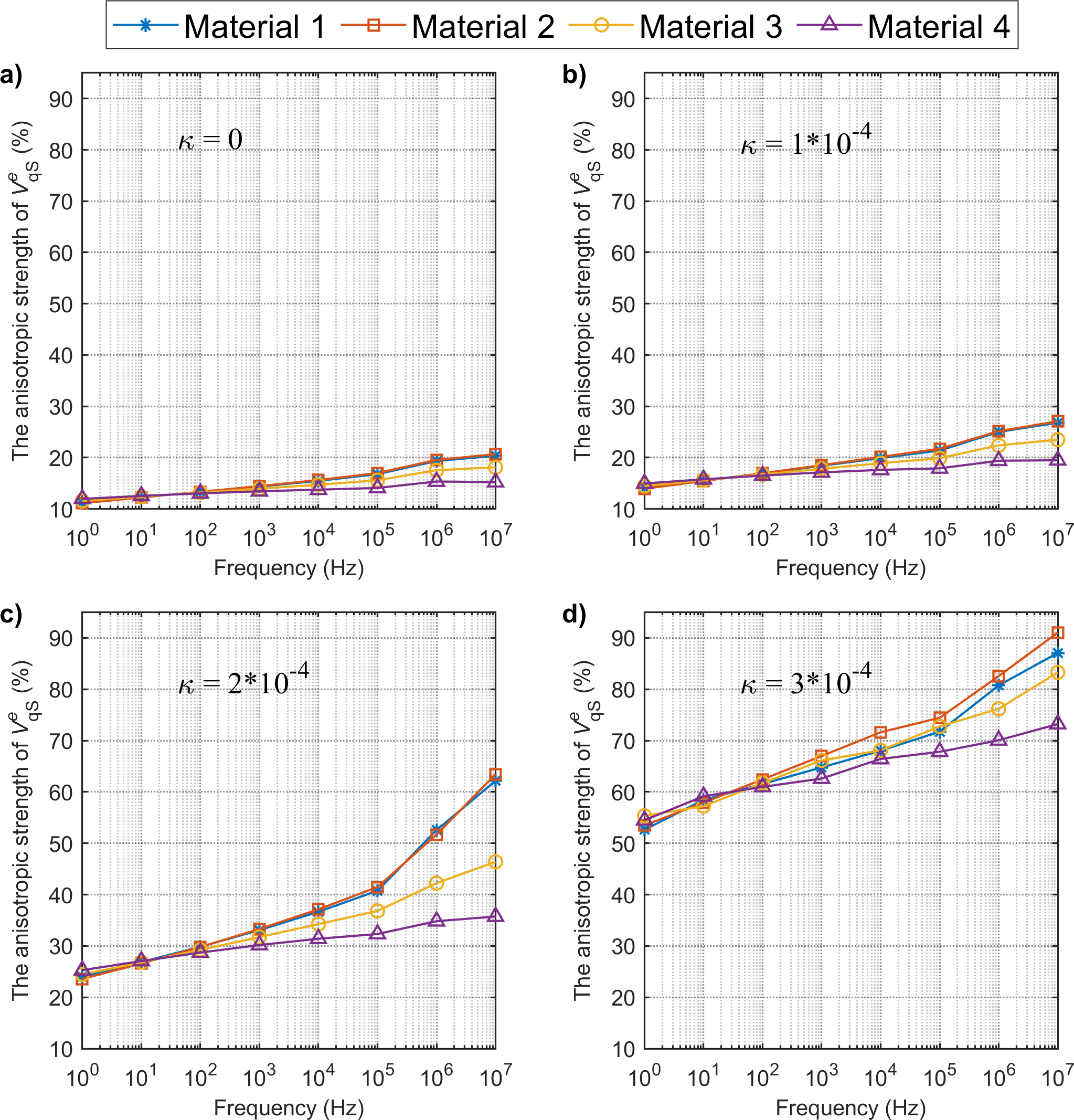}
\caption{Similar with Figure \ref{fig8}, but for the qS wave.}
\label{fig9}
\end{figure}

\begin{figure}
\centering
\includegraphics[width=0.66\textwidth]{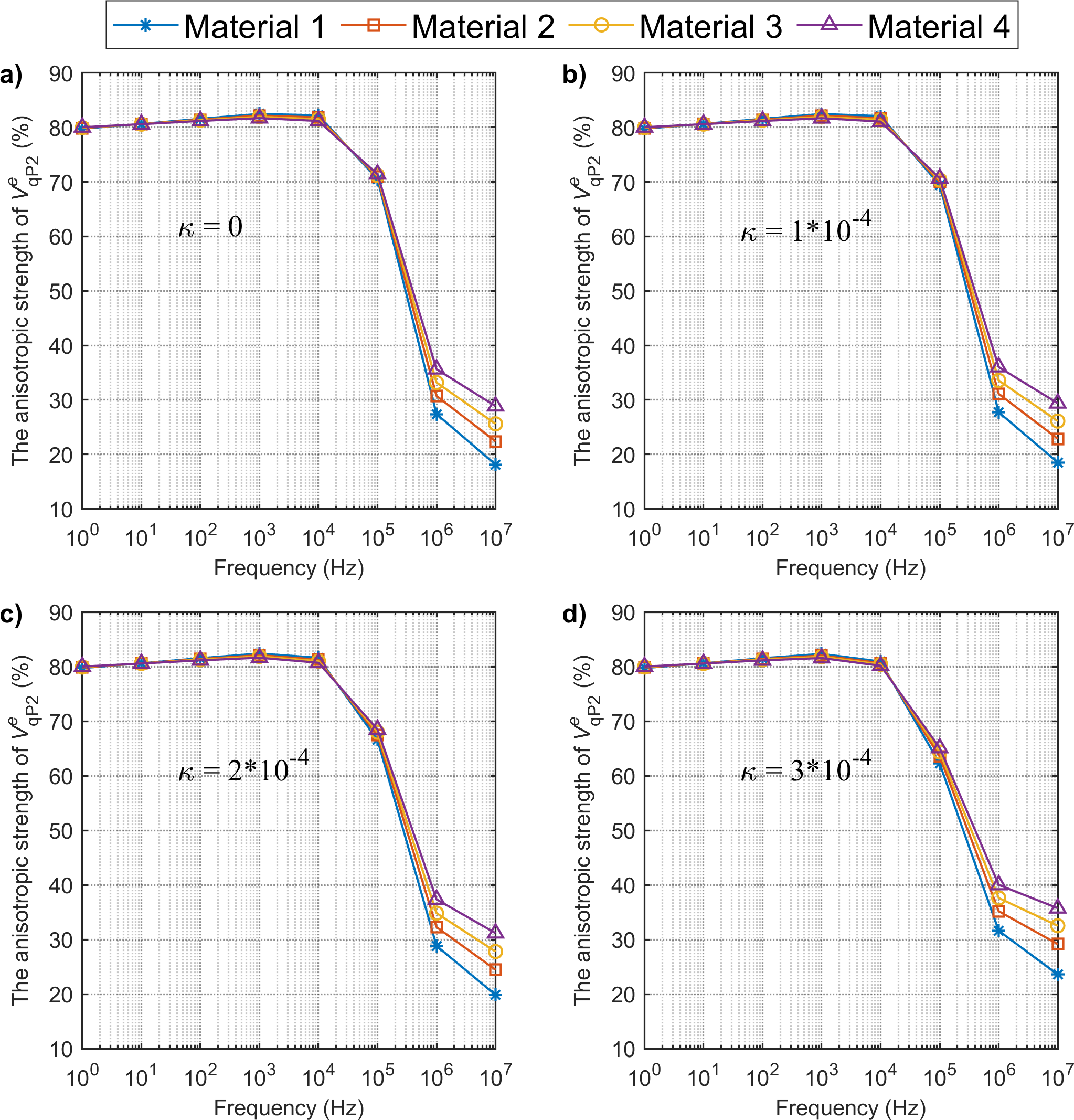}
\caption{Similar with Figure \ref{fig8}, but for the qP2 wave.}
\label{fig10}
\end{figure}

\begin{figure}
\centering
\includegraphics[width=0.66\textwidth]{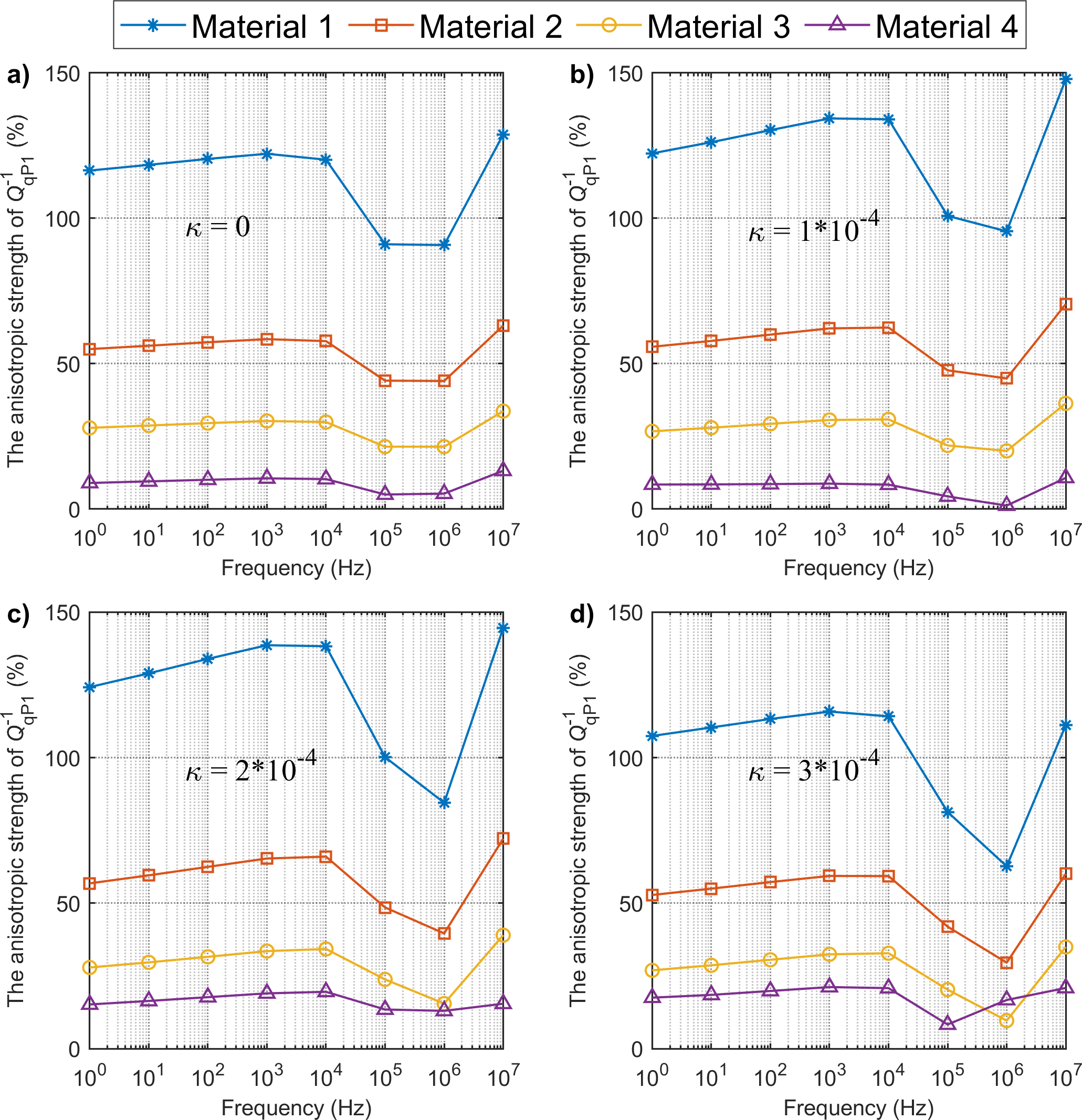}
\caption{The anisotropic factor curves of the qP1-wave dissipation factor, where (a), (b), (c), and (d), correspond to inhomogeneity parameter $\kappa =0$, $1*{{10}^{-4}}$, $2*{{10}^{-4}}$, and $3*{{10}^{-4}}$, respectively. The anisotropic factors are calculated at frequencies $1$, ${{10}^{1}}$, ${{10}^{2}}$, ${{10}^{3}}$, ${{10}^{4}}$, ${{10}^{5}}$, ${{10}^{6}}$, and ${{10}^{7}}$, respectively.}
\label{fig11}
\end{figure}

\begin{figure}
\centering
\includegraphics[width=0.66\textwidth]{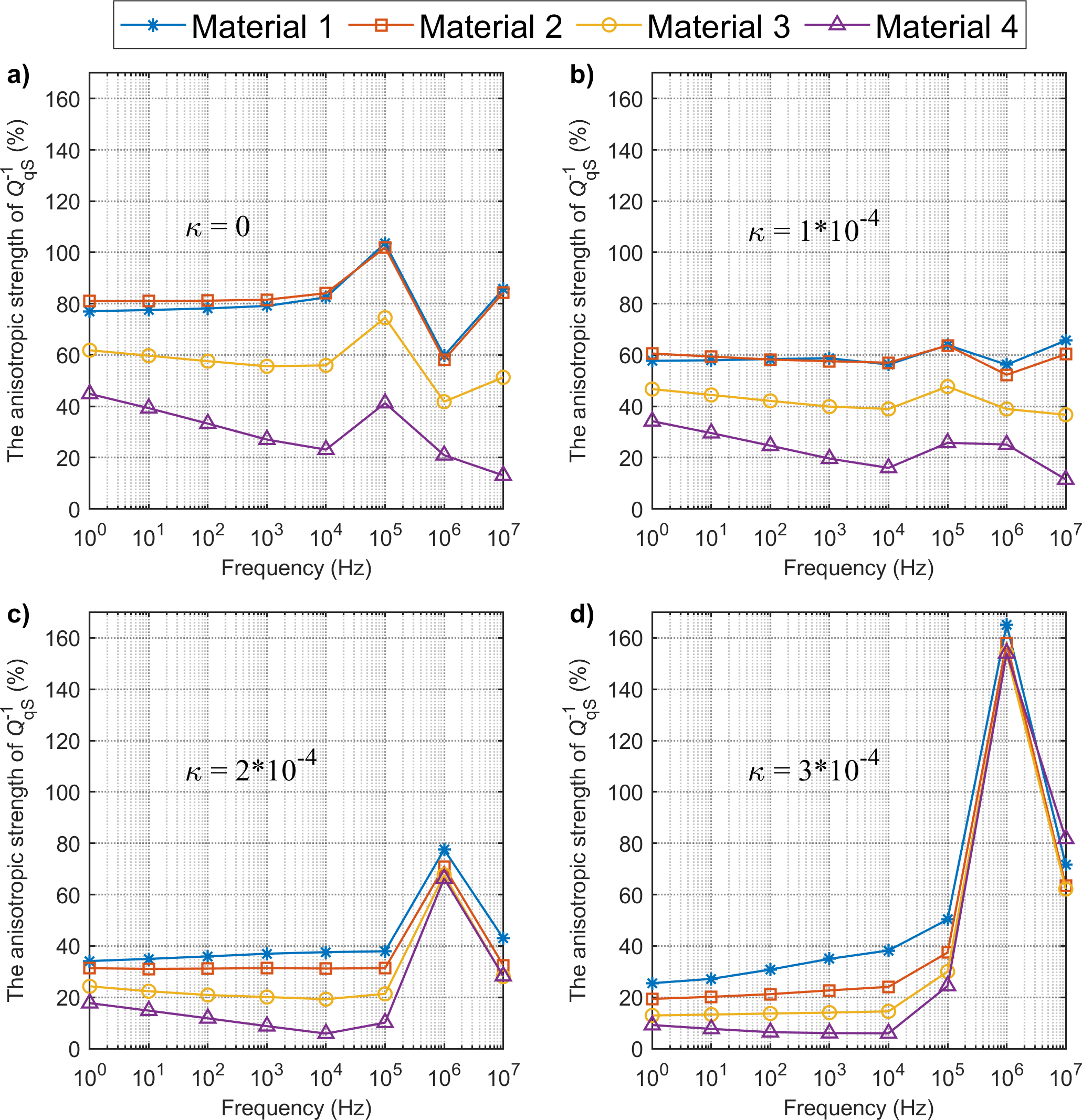}
\caption{Similar with Figure \ref{fig11}, but for the qS wave.}
\label{fig12}
\end{figure}

\begin{figure}
\centering
\includegraphics[width=0.66\textwidth]{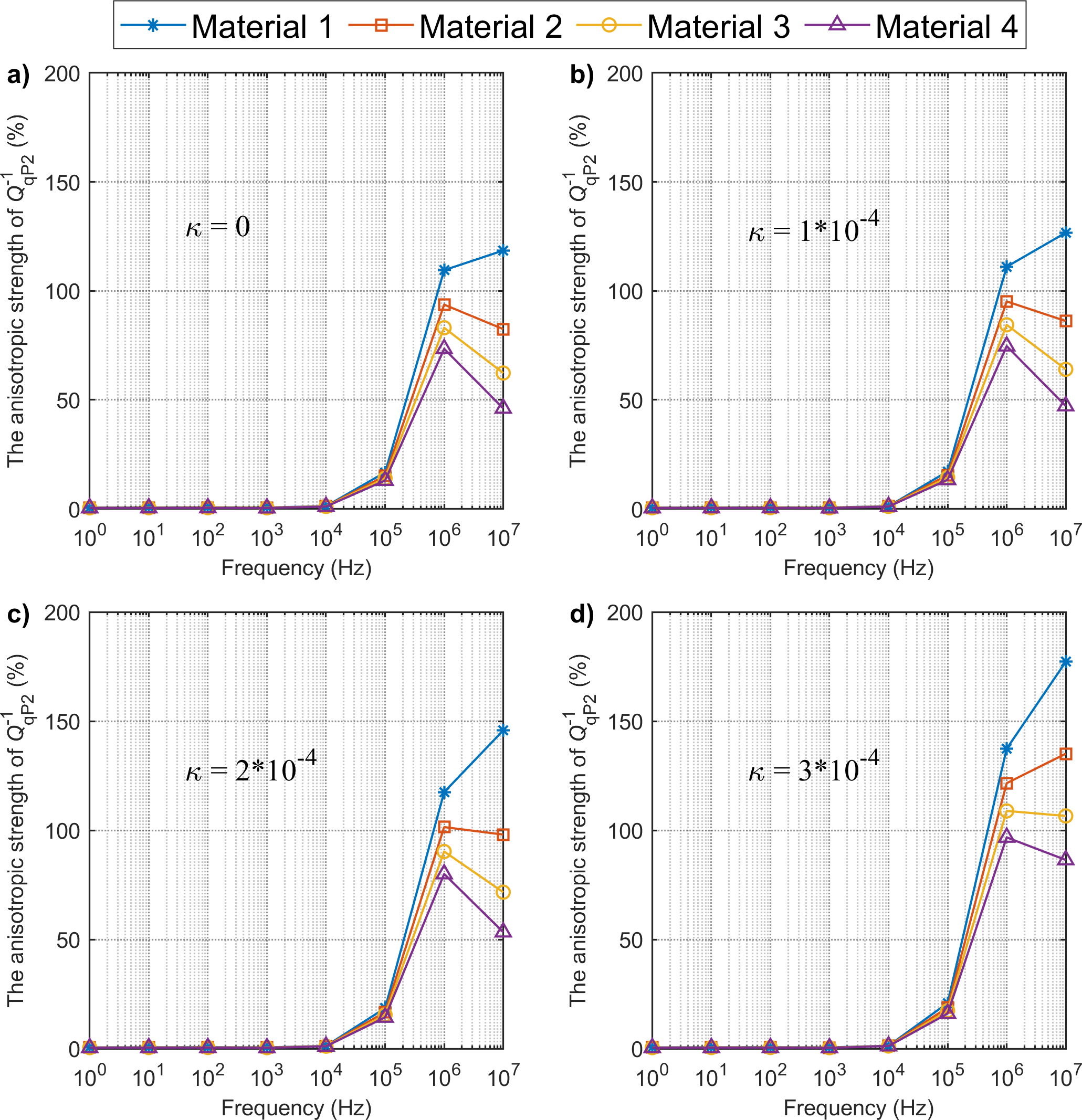}
\caption{Similar with Figure \ref{fig11}, but for the qP2 wave.}
\label{fig13}
\end{figure}

\subsection{Effect of attenuation anisotropy and wave inhomogeneity}
To further study the propagation of seismic waves in the proposed model, we analyze how attenuation anisotropy and wave inhomogeneity affect the wave behavior in a wide frequency range. To fully consider a broad range of attenuation anisotropic strength, we present four attenuative VTI porous materials, which are listed in Table \ref{tab2}. Note that here we invoke the Thomsen-style parameters ${{\epsilon }_{Q}}$ and ${{\delta }_{Q}}$ proposed by \cite{zhu2006plane} to better quantify the strength of attenuation anisotropy. Other properties of the four materials are the same and are defined in Table \ref{tab3}. As shown in Table \ref{tab2}, the strength of attenuation anisotropy gradually decreases from materials 1 to 3, and material 4 is attenuation isotropy. Table \ref{tab3} shows that the velocities and the fluid flows of the four materials are anisotropic. In the subsection, We emphasize that all dissipation factors are deduced by using the definition of $Q_{2}^{-1}$.

\begin{table}
  \centering
  \caption{The strengths of attenuation anisotropy for four attenuative VTI porous materials.}
  \label{tab2}
  \begin{tabular}{@{} cc  cc  cc  cc @{}}
    \toprule
    \multicolumn{2}{c}{Material 1} & \multicolumn{2}{c}{Material 2}
    & \multicolumn{2}{c}{Material 3} & \multicolumn{2}{c}{Material 4} \\
    \cmidrule(lr){1-2} \cmidrule(lr){3-4}
    \cmidrule(lr){5-6} \cmidrule(lr){7-8}
    $\epsilon_{Q}$ & $\delta_{Q}$ & $\epsilon_{Q}$ & $\delta_{Q}$
    & $\epsilon_{Q}$ & $\delta_{Q}$ & $\epsilon_{Q}$ & $\delta_{Q}$ \\
    \midrule
    -0.5 & -0.8  & -0.3  & -0.6  & -0.15 & -0.3 & 0 & 0 \\
    \bottomrule
  \end{tabular}
\end{table}

\begin{table*}
  \centering
  \caption{The common properties of four attenuative VTI porous materials. The units of $C_{IJ}$ and $K_{ii}$ are in GPa and mD, respectively.}
  \label{tab3}
  \begin{tabular}{@{} *{12}{c} @{}}
    \toprule
    $C_{11}$ & $C_{12}$ & $C_{13}$ & $C_{33}$ & $C_{55}$
    & $Q^{s}_{33}$ & $Q^{s}_{55}$ & $Q^{f}_{3}$
    & $K_{11}$ & $K_{33}$ & $T_{1}$ & $T_{3}$ \\
    \midrule
    32 & 16 & 12 & 23 & 6 & 25 & 60 & 20 & 300 & 100 & 2.2 & 3.2 \\
    \bottomrule
  \end{tabular}
\end{table*}

Here, we first clarify a concept that attenuation anisotropy is not identical to anelastic. Generally, anelastic is proportional to velocity dispersion and energy attenuation. However, the increasing strength of attenuation anisotropy does not necessarily lead to more notable dispersion and energy attenuation. To illustrate the difference between them, the energy velocity and the dissipation factor curves of the qP1 wave at the inhomogeneity parameter $\kappa =1*{{10}^{-4}}$ are shown in Figure \ref{fig7}, where the propagation angle is ${\pi }/{4}\;$. We can find that compared with materials 2, 3, and 4, the frequency dependence of the energy velocity is the smallest in material 1 with the strongest attenuation anisotropy strength, and the energy attenuation is the weakest. Therefore, we cannot confuse the strengths of the anelastic and the attenuation anisotropy, which have different effects on wave behavior.

Next, to measure the anisotropic strengths (i.e., directional dependence) of the energy velocities and the dissipation factors of the qP1, qS, and qP2 waves in the porous materials with different attenuation anisotropy strengths, we define the anisotropy factors as follows
\begin{equation}\tag{57.1}\label{eq57.1}
\Xi _{\text{qP1,qS,qP2}}^{V}=100\times \frac{\max \left[ V_{\text{qP1,qS,qP2}}^{e}\left( \theta  \right) \right]-\min \left[ V_{\text{qP1,qS,qP2}}^{e}\left( \theta  \right) \right]}{\min \left[ V_{\text{qP1,qS,qP2}}^{e}\left( \theta  \right) \right]},     
\end{equation}
\begin{equation}\tag{57.2}\label{eq57.2}
\Xi _{\text{qP1,qS,qP2}}^{Q}=100\times \frac{\max \left[ Q_{\text{qP1,qS,qP2}}^{-1}\left( \theta  \right) \right]-\min \left[ Q_{\text{qP1,qS,qP2}}^{-1}\left( \theta  \right) \right]}{\min \left[ Q_{\text{qP1,qS,qP2}}^{-1}\left( \theta  \right) \right]},
\end{equation}
where $\Xi _{\text{qP1}}^{V}$, $\Xi _{\text{qS}}^{V}$, and $\Xi _{\text{qP2}}^{V}$ represent the anisotropic strengths of the qP1-, qS-, and qP2-wave energy velocities, respectively, and $\Xi _{\text{qP1}}^{Q}$, $\Xi _{\text{qS}}^{Q}$, and $\Xi _{\text{qP2}}^{Q}$ denote the anisotropic strengths of the dissipation factors of the qP1, qS, and qP2 waves, respectively. It can be well understood that the greater the value of the anisotropy factors, the stronger the anisotropy strengths of the energy velocities and the dissipation factors.

In the four materials, the anisotropic factor curves of the energy velocities for the qP1, qS, and qP2 waves as functions of frequency are shown in Figures \ref{fig8}, \ref{fig9}, and \ref{fig10}, respectively. Figures \ref{fig11}, \ref{fig12}, and \ref{fig13} present the corresponding anisotropic factor curves of the dissipation factors. In these figures, panels a, b, c, and d correspond to inhomogeneity parameter $\kappa =0$, $1*{{10}^{-4}}$, $2*{{10}^{-4}}$, and $3*{{10}^{-4}}$, respectively. For each panel, we calculate the anisotropy factors at frequencies $1$, ${{10}^{1}}$, ${{10}^{2}}$, ${{10}^{3}}$, ${{10}^{4}}$, ${{10}^{5}}$, ${{10}^{6}}$, and ${{10}^{7}}$, respectively.

From Figures \ref{fig8}, \ref{fig9}, and \ref{fig10}, we can observe that:
\begin{itemize}
    \item For the qP1 wave, attenuation anisotropy causes a more pronounced frequency-dependent behavior of parameter $\Xi _{\text{qP1}}^{V}$, indicating that the velocity anisotropy of the qP1 wave becomes strongly frequency-dependent in this scenario. In contrast, the influence of attenuative isotropic porous materials on parameter $\Xi _{\text{qP1}}^{V}$ is relatively subtle. This means that the velocity anisotropy of the qP1 wave in attenuative isotropic porous media exhibits only slight frequency dependence. However, the frequency dependence of the qP1 wave velocity anisotropy (represented by $\Xi _{\text{qP1}}^{V}$) does not increase proportionally with attenuation anisotropy strength; rather, it is influenced by the inhomogeneity parameter $\kappa$.
    \item For the qS wave, the $\Xi _{\text{qS}}^{V}$ curves for all four material types nearly overlap within the exploration frequency band ($<10^3$ Hz). This suggests that varying strengths of attenuation anisotropy have minimal impact on qS-wave velocity anisotropy in the exploration frequency range. Nonetheless, a similar phenomenon to that of the qP1 wave is observed: the introduction of attenuation anisotropy also results in a stronger frequency dependence of qS-wave velocity anisotropy.
    \item Regarding the qP2 wave, $\Xi _{\text{qP2}}^{V}$ is barely influenced by attenuation anisotropy in the frequency range up to about Biot relaxation frequency. While the frequencies exceed the Biot relaxation frequency, velocity anisotropy decreases rapidly, and its decrease is positively correlated with the attenuation anisotropy.
    \item The shared influence of the inhomogeneity parameter $\kappa$ on the velocity anisotropy of all three wave modes. Specifically, regardless of whether the porous material exhibits attenuation anisotropy or attenuation isotropy, increasing the inhomogeneity parameter $\kappa$ elevates the anisotropy strength. This suggests that an increase in the magnitude of inhomogeneity parameter $\kappa$ leads to a stronger velocity anisotropy across the qP1, qS, and qP2 waves.
\end{itemize}

From Figures \ref{fig11}, \ref{fig12}, and \ref{fig13}, we can see that:
\begin{itemize}
    \item For the qP1 wave, at frequencies lower than ${{10}^{5}}$ Hz, the directional dependence of the dissipation factor becomes more pronounced with increasing attenuation anisotropy. Around the Biot relaxation frequency, the directional dependence of the dissipation factor is markedly reduced in the attenuative anisotropic porous materials. The magnitude of this decrease is proportional to the attenuation anisotropic strength of the materials, which may lead to the direction dependence of the qP1-wave dissipation factor in the attenuative anisotropic porous material (i.e., material 3 in Figures \ref{fig11}d) is weaker than that of in the attenuative isotropic porous material 4.
    \item For the qS wave, incorporating attenuation anisotropy results in a more pronounced directional dependence of the dissipation factor. However, the stronger attenuation anisotropy does not necessarily mean the greater the directional dependence of the dissipation factor, which is related to the inhomogeneity parameter $\kappa$. Compared with the qP1-wave behavior near the Biot relaxation frequency, the directional dependence of the qS-wave dissipation factor exhibits a significant increase, especially for a surprising amplitude change shown in Figure \ref{fig12}d.
    \item For the qP2 wave, at frequencies below ${{10}^{4}}$ Hz, its dissipation factor does not exhibit directional dependence. This is because the qP2 wave is a diffusive wave, and its dissipation factor under the definition of $Q_{2}^{-1}$ remains approximately 2 in every propagation direction within this frequency band (see Figure \ref{fig6}c). As the frequency increases, the attenuation anisotropy gradually affects the behavior of the qP2-wave dissipation factor, especially in the frequency exceeding the Biot relaxation frequency, the qP2-wave dissipation factor shows a very strong directional dependence. More interestingly, the directional dependence of the qP2-wave dissipation factor is more pronounced in attenuative isotropic material than in attenuative anisotropic materials.
    \item As observed in the influence of the inhomogeneity parameter on velocity anisotropy, we likewise see that the inhomogeneity parameter $\kappa$ affects the directional dependence of the dissipation factor. The coupled effects of inhomogeneity and attenuation anisotropy govern the directional dependence of the dissipation factors for all three wave modes.
\end{itemize}

The numerical results demonstrate that attenuation anisotropy exerts markedly different impacts on the three wave modes across the explored frequency band. Meanwhile, the wave inhomogeneity inferences both velocity and attenuation anisotropies, underscoring the coupled role of material heterogeneity and attenuation anisotropy in controlling wave propagation behavior.  

\begin{figure}
\centering
\includegraphics[width=0.66\textwidth]{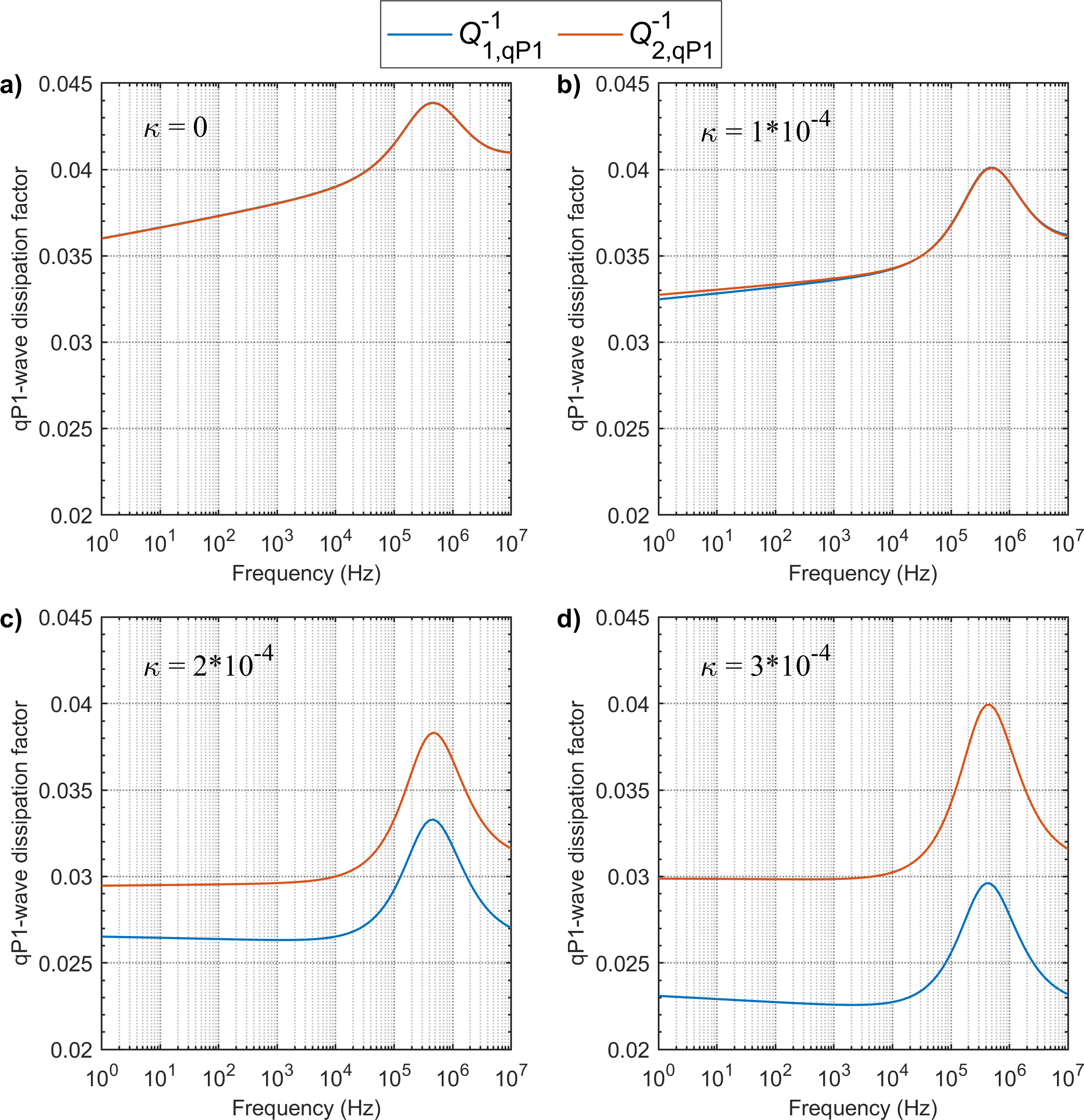}
\caption{Comparison between $Q_{\text{1,qP}1}^{-1}$ (blue line) and $Q_{\text{2,qP}1}^{-1}$ (orange line) of the qP1 wave for the attenuative VTI porous material (material 2 shown in Table \ref{tab2}) at different inhomogeneity parameter $\kappa$, where the panels (a), (b), (c), and (d) correspond to the values of $\kappa =0$, $1*{{10}^{-4}}$, $2*{{10}^{-4}}$, and $3*{{10}^{-4}}$, respectively.}
\label{fig14}
\end{figure}

\begin{figure}
\centering
\includegraphics[width=0.66\textwidth]{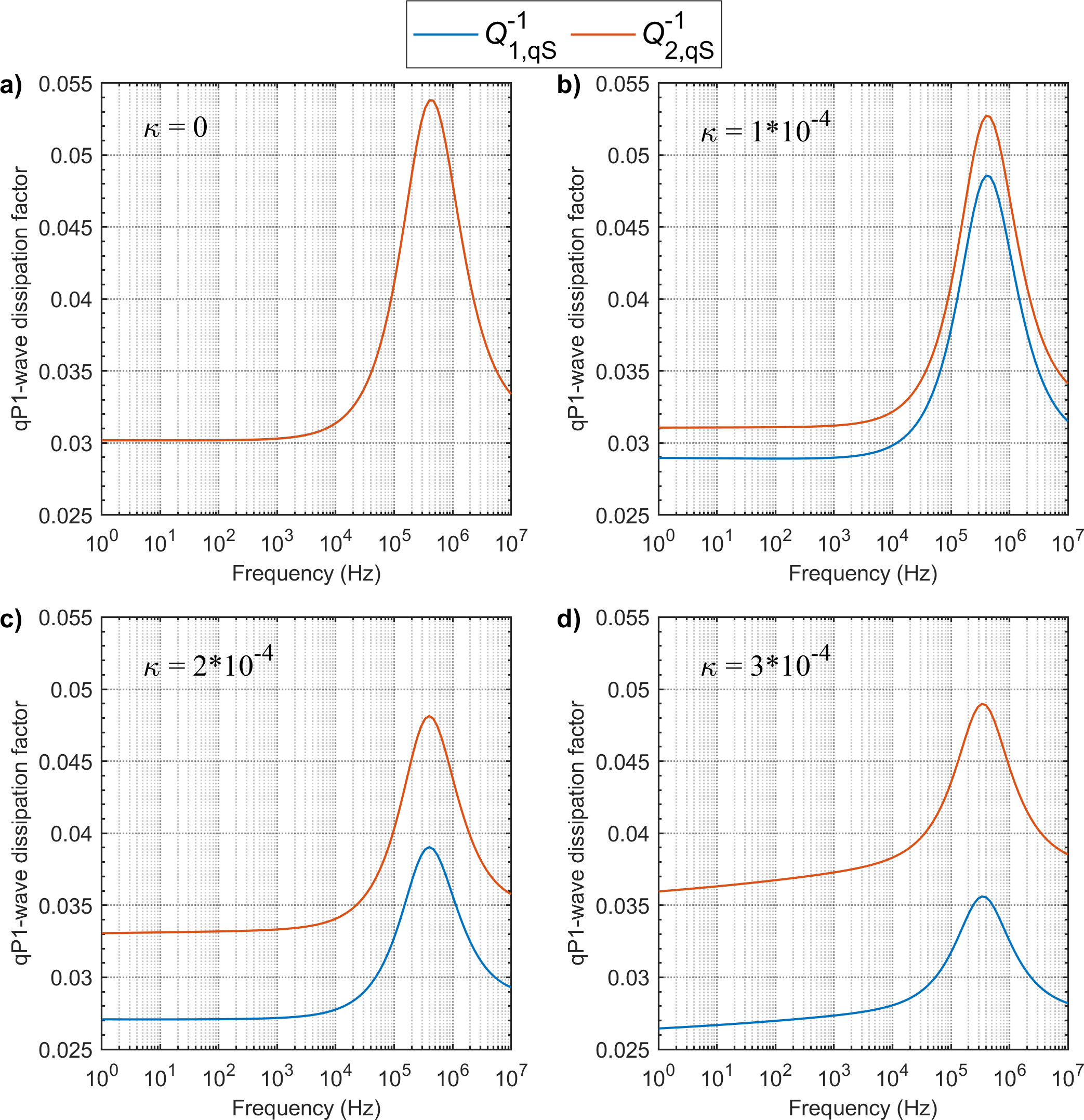}
\caption{Similar with Figure \ref{fig14}, but for the qS wave.}
\label{fig15}
\end{figure}

\begin{figure}
\centering
\includegraphics[width=0.66\textwidth]{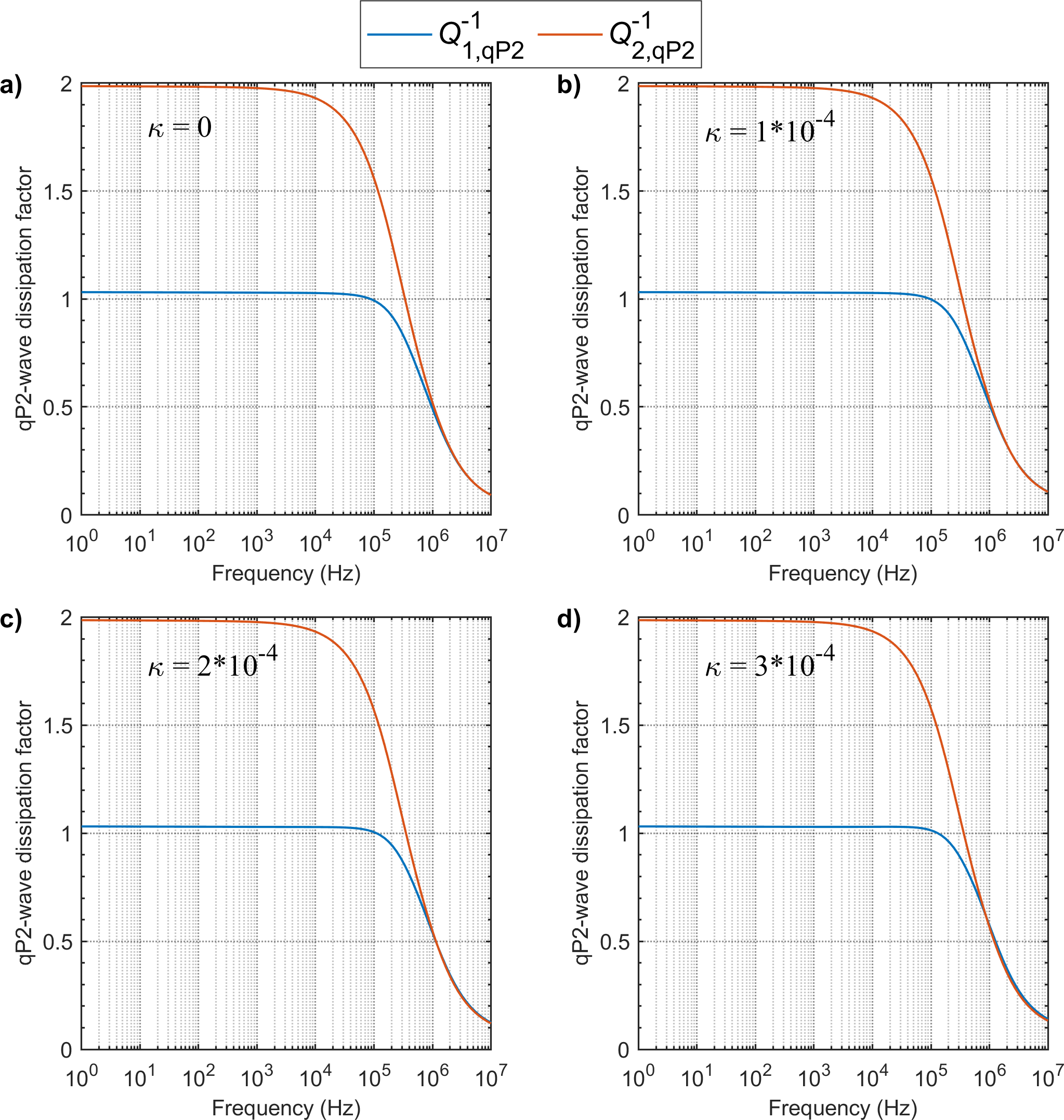}
\caption{Similar with Figure \ref{fig14}, but for the qP2 wave.}
\label{fig16}
\end{figure}

\subsection{Comparison of two definitions of dissipation factor}
As stated in Section \ref{sec:theorysub5}, we derived two explicit formulations for the dissipation factor in attenuative anisotropic media, based on the distinct definitions $Q_{1}^{-1}$ and $Q_{2}^{-1}$. Here, we present a numerical example to illustrate their divergence. We again consider an attenuative anisotropic medium whose elastic parameters are listed in Tables \ref{tab1} and \ref{tab3}, with attenuation anisotropy parameters taken from Material 2 in Table \ref{tab2}. We also examine four degrees of wave inhomogeneity, corresponding to inhomogeneity parameters $\kappa =0$, $1*{{10}^{-4}}$, $2*{{10}^{-4}}$, and $3*{{10}^{-4}}$, respectively, and set the wave propagation angle to ${\pi }/{4}\;$

The comparison between $Q_{\text{1,qP}1}^{-1}$ and $Q_{\text{2,qP}1}^{-1}$ of the qP1 wave is shown in Figure \ref{fig14}. The difference between $Q_{\text{1,qP}1}^{-1}$ and $Q_{\text{2,qP}1}^{-1}$ becomes remarkable with increasing absolute value of $\kappa $. As displayed in Figure \ref{fig15}, the same phenomenon occurs between $Q_{\text{1,qS}}^{-1}$ and $Q_{\text{2,qS}}^{-1}$. In particular, Figures \ref{fig14}a and \ref{fig15}a indicate that the dissipation factor curves under definitions $Q_{\text{1}}^{-1}$ and $Q_{\text{2}}^{-1}$ coincide for the homogeneous qP1 and qS waves. Following the two definition expressions (\ref{eq53}) and (\ref{eq54}) under the condition of the homogeneous waves, we can find that the denominator term $\operatorname{Im}\left( s \right){\operatorname{Im}\left[ s{{\overline{\boldsymbol{V}}}^{\operatorname{T}}}\cdot \mathrm{ }\!\!\Gamma\!\!\text{ }\cdot \boldsymbol{V} \right]}/{\operatorname{Re}\left[ s{{\overline{\boldsymbol{V}}}^{\operatorname{T}}}\cdot \mathrm{ }\!\!\Gamma\!\!\text{ }\cdot \boldsymbol{V} \right]}\;$ in the formula $Q_{1\text{h}}^{-1}$ of the qP1 and qS waves tends to be zero. On the other hand, according to the definition expressions (\ref{eq45}) and (\ref{eq46}), we can also conclude that the average strain energy density ${{E}_{s}}$ is equal to the average kinetic energy density ${{E}_{v}}$ in the case of homogeneous wave, which means that the term $\operatorname{Im}\left[ {{\overline{\boldsymbol{V}}}^{\operatorname{T}}}\cdot \operatorname{Im}\left( \boldsymbol{S} \right)\cdot \boldsymbol{R}\cdot {{\boldsymbol{S}}^{\text{T}}}\cdot \boldsymbol{V} \right]$ is identical to zero due to equation (\ref{eq50}).

Figure \ref{fig16} compares the two definitions of the qP2‐wave dissipation factor,
$Q^{-1}_{1,\mathrm{qP2}}$ and $Q^{-1}_{2,\mathrm{qP2}}$ over a range of
inhomogeneity parameters $\kappa$. We can see that under the two distinct definitions, the qP2-wave dissipation factors exhibit a pronounced divergence in the low frequency band ( e.g., $<10^{4}$ Hz), yielding nearly constant values of approximately 1 and 2, respectively. Beyond the relaxation frequency, however, the dissipation factors computed from both definitions converge to almost identical values. In terms of the definition expressions (\ref{eq45}) and (\ref{eq46}), these observations reflect the energy‐partitioning characteristics of the qP2 wave: 
\begin{itemize}
    \item In the low‐frequency range, the qP2 wave is a dissipative wave with a very slow velocity and, thus, its average kinetic energy density \(E_v\) is negligible compared to its average strain‐energy density \(E_s\) (i.e.\ \(E_v/E_s \to 0\)). As a result, most of the seismic wave energy is stored in the strain energy.
    \item Once the frequency exceeds the relaxation point, \(E_v\) and \(E_s\) approach equality (\(E_v \approx E_s\)), yielding an approximately equal split of seismic energy between kinetic and strain forms.
\end{itemize}

%% file: Sections/Discussion.tex
\section{\textbf{Discussion}}\label{sec:discussion}
The traditional Biot model significantly underestimates dispersion and attenuation across a wide frequency range. Our proposed model effectively addresses these limitations by considering the coupled relaxed skeleton (RS) or the relaxed fluid diffusion (RF) effects, resulting in markedly stronger dispersion and attenuation, particularly for the fast quasi-compressional (qP1) and quasi-shear (qS) waves relevant to seismic exploration. The comparative analysis further demonstrated that while the RS mechanism predominantly influences the dispersion and attenuation of qP1 and qS waves, the RF mechanism notably impacts slow quasi-compressional (qP2) wave behavior, particularly at frequencies exceeding the Biot relaxation frequency. These findings highlight the mechanism-specific roles in controlling the complex dynamic responses of different wave modes.

Our results also clarify the conceptual distinction between anelasticity and attenuation anisotropy. While anelasticity directly relates to overall energy dissipation and dispersion, attenuation anisotropy specifically captures directional variations in these properties. Interestingly, increased attenuation anisotropy strength does not necessarily translate into greater overall attenuation or dispersion, as demonstrated in our comparative tests involving varying strengths of attenuation anisotropy. Another critical aspect revealed in this study is the role of wave inhomogeneity, which significantly affects both velocity and attenuation anisotropy, underscoring a coupled relationship with material attenuation properties. Such coupling emphasizes the complexity of wave propagation phenomena, where directional and frequency-dependent behaviors become prominent. Moreover, our comparative assessment of two commonly used definitions for dissipation factors ($Q_1^{-1}$ and $Q_2^{-1}$) indicates considerable divergence for inhomogeneous waves, especially pronounced for qP2 waves at lower frequencies. This discrepancy stems from differing assumptions underlying these definitions about the partitioning of seismic energy into kinetic and strain components. 

Overall, this work provides a new model for studying wave propagation in poro-viscoelastic media with the anisotropy of attenuation and velocity. Since our derivation starts from the general anisotropy, our proposed theory can be well degraded to some common anisotropy, such as monoclinic, orthorhombic, and transverse isotropy. Moreover, we use a more general inhomogeneous plane wave theory instead of the conventional homogeneous plane wave theory to present the concise expressions of measurable quantities and fundamental relations. Compared with the traditional expressions of measurable quantities, our expressions can more accurately predict the dispersion and attenuation in attenuative porous media, which has practical purpose for the comprehensively understanding seismic waves propagation in complex media combining viscoelastic, anisotropic, and poroelastic properties.

%% file: Sections/Conclusions.tex
\section{\textbf{Conclusions}}
We proposed a complete model to describe attenuative anisotropic porous media. This completeness is mainly reflected in two aspects. Firstly, the proposed model has a complete anisotropy, that is, the poro-viscoelastic model includes both the anisotropy of velocity and attenuation; Secondly, the complete relaxed skeleton (RS) or the relaxed fluid diffusion (RF) mechanisms are unified, which is considered simultaneously in the constitutive relations, the dynamic Darcy’s law, and the fluid pressure. Moreover, based on the method of complex slowness vector, we derived the Christoffel and energy balance equations by using inhomogeneous plane wave analysis, while the phase velocities and complex slowness of different wave modes are obtained by solving an eighth-degree algebraic equation. To measure the dispersion and attenuation caused by viscodynamic effects and the RS and RF mechanisms, the concise expressions of energy velocity and dissipation factor under the distinct definitions are derived from energy consideration. These explicit expressions make it easy to simulate inhomogeneous plane waves propagation in any specified direction.

The characteristics of inhomogeneous plane waves propagating in 2-D unbounded attenuative vertical transversely isotropic porous materials are investigated numerically. Our findings demonstrated that incorporating RS and RF mechanisms leads to an improved dispersion and attenuation compared to traditional Biot-based models. Additionally, attenuation anisotropy distinctly affects wave propagation behavior, introducing frequency-dependent directional variations in dispersion and attenuation. Also, wave inhomogeneity notably influences velocity and attenuation anisotropies, underscoring its essential role in realistic seismic modeling. Furthermore, significant discrepancies between two widely-used dissipation factor definitions arise in inhomogeneous wave scenarios, emphasizing the need for careful consideration based on wave modes and media characteristics.

%% file: Sections/CodeAvailability.tex
\section{\textbf{Data and code availability statement}}
Data and code associated with this research are available and can be obtained by contacting the corresponding author.

%% file: Sections/Appendix.tex
\section{Appendix}
\subsection{The energy balance equation in attenuative anisotropic porous media}\label{appendA}
In this appendix, we derive the energy balance equation in attenuative anisotropic porous media based on the approach of the complex slowness. Here, we assume that the source is absent.
The dot product of the complex conjugate of the equation (\ref{eq26}) with $-{{\boldsymbol{T}}^{\operatorname{T}}}$ yields
\begin{equation}\tag{A1}\label{eqA1}
-{{\boldsymbol{T}}^{\operatorname{T}}}\cdot {{\boldsymbol{D}}^{\operatorname{T}}}\cdot \overline{\boldsymbol{V}}=\operatorname{i}\omega {{\boldsymbol{T}}^{\operatorname{T}}}\cdot \overline{\boldsymbol{\Pi }}.
\end{equation}

The dot product of ${{\overline{\boldsymbol{V}}}^{\operatorname{T}}}$ with the equation (\ref{eq27}) becomes
\begin{equation}\tag{A2}\label{eqA2}
{{\overline{\boldsymbol{V}}}^{\operatorname{T}}}\cdot \boldsymbol{D}\cdot \boldsymbol{T}=-\text{i}\omega {{\overline{\boldsymbol{V}}}^{\operatorname{T}}}\cdot \boldsymbol{G}\cdot \boldsymbol{V}.
\end{equation}

The equation (\ref{eqA1}) minus (\ref{eqA2}) gives
\begin{equation}\tag{A3}\label{eqA3}
-{{\boldsymbol{T}}^{\operatorname{T}}}\cdot {{\boldsymbol{D}}^{\operatorname{T}}}\cdot \overline{\boldsymbol{V}}-{{\overline{\boldsymbol{V}}}^{\operatorname{T}}}\cdot \boldsymbol{D}\cdot \boldsymbol{T}=\operatorname{i}\omega {{\boldsymbol{T}}^{\operatorname{T}}}\cdot \overline{\boldsymbol{\Pi }}\text{+i}\omega {{\overline{\boldsymbol{V}}}^{\operatorname{T}}}\cdot \boldsymbol{G}\cdot \boldsymbol{V}.
\end{equation}

The left-hand side of equation (\ref{eqA3}) can be simplified to
\begin{equation}\tag{A4}\label{eqA4}
-{{\boldsymbol{T}}^{\operatorname{T}}}\cdot {{\boldsymbol{D}}^{\operatorname{T}}}\cdot \overline{\boldsymbol{V}}-{{\overline{\boldsymbol{V}}}^{\operatorname{T}}}\cdot \boldsymbol{D}\cdot \boldsymbol{T}=2\nabla \cdot \boldsymbol{p},
\end{equation}
where the quantity $\boldsymbol{p}$ is the complex Umov–Poynting vector presented as follows:
\begin{equation}\tag{A5}\label{eqA5}
\boldsymbol{p}=-\frac{1}{2}\left[ \begin{matrix}
   {{\sigma }_{11}} & {{\sigma }_{12}} & {{\sigma }_{13}} & -P & 0 & 0  \\
   {{\sigma }_{12}} & {{\sigma }_{22}} & {{\sigma }_{23}} & 0 & -P & 0  \\
   {{\sigma }_{13}} & {{\sigma }_{23}} & {{\sigma }_{33}} & 0 & 0 & -P  \\
\end{matrix} \right]\cdot \overline{\boldsymbol{V}}.
\end{equation}

With equation (\ref{eqA4}), the equation (\ref{eqA3}) becomes
\begin{equation}\tag{A6}\label{eqA6}
\begin{split}
\nabla \cdot \boldsymbol{p}=\, &2\text{i}\omega \left\{ \frac{1}{4}\operatorname{Re}\left[ {{\boldsymbol{T}}^{\operatorname{T}}}\cdot \overline{\boldsymbol{\Pi }} \right]+\frac{1}{4}\operatorname{Re}\left[ {{\overline{\boldsymbol{V}}}^{\operatorname{T}}}\cdot \boldsymbol{G}\cdot \boldsymbol{V} \right] \right\} \\
&-\omega \left\{ \frac{1}{2}\operatorname{Im}\left[ {{\boldsymbol{T}}^{\operatorname{T}}}\cdot \overline{\boldsymbol{\Pi }} \right]+\frac{1}{2}\operatorname{Im}\left[ {{\overline{\boldsymbol{V}}}^{\operatorname{T}}}\cdot \boldsymbol{G}\cdot \boldsymbol{V} \right] \right\},
\end{split}
\end{equation}
Each term on the right-hand side of equation (\ref{eqA6}) has a precise physical meaning in terms of an average time. Among them, the average strain and kinetic energy densities are given by \citep{carcione2014wave}
\begin{equation}\tag{A7}\label{eqA7}
{{E}_{s}}=\frac{1}{4}\operatorname{Re}\left[ {{\boldsymbol{T}}^{\operatorname{T}}}\cdot \overline{\boldsymbol{\Pi }} \right],
\end{equation}
and
\begin{equation}\tag{A8}\label{eqA8}
{{E}_{v}}=\frac{1}{4}\operatorname{Re}\left[ {{\overline{\boldsymbol{V}}}^{\operatorname{T}}}\cdot \boldsymbol{G}\cdot \boldsymbol{V} \right],
\end{equation}
respectively, and the average dissipated strain and kinetic energy densities can be expressed as
\begin{equation}\tag{A9}\label{eqA9}
{{E}_{ds}}=-\frac{1}{2}\operatorname{Im}\left[ {{\boldsymbol{T}}^{\operatorname{T}}}\cdot \overline{\boldsymbol{\Pi }} \right],
\end{equation}
and
\begin{equation}\tag{A10}\label{eqA10}
{{E}_{dv}}=\frac{1}{2}\operatorname{Im}\left[ {{\overline{\boldsymbol{V}}}^{\operatorname{T}}}\cdot \boldsymbol{G}\cdot \boldsymbol{V} \right],
\end{equation}
respectively. 

The sum of equations (\ref{eqA7}) and (\ref{eqA8}) is called the average stored energy density as
\begin{equation}\tag{A11}\label{eqA11}
{{E}_{a}}={{E}_{s}}+{{E}_{v}}=\frac{1}{4}\operatorname{Re}\left[ {{\boldsymbol{T}}^{\operatorname{T}}}\cdot \overline{\boldsymbol{\Pi }}+{{\overline{\boldsymbol{V}}}^{\operatorname{T}}}\cdot \boldsymbol{G}\cdot \boldsymbol{V} \right].
\end{equation}

Equation (\ref{eqA9}) plus equation (\ref{eqA10}) equals the average dissipated energy density as
\begin{equation}\tag{A12}\label{eqA12}
{{E}_{d}}={{E}_{dv}}+{{E}_{ds}}=\frac{1}{2}\operatorname{Im}\left[ {{\overline{\boldsymbol{V}}}^{\operatorname{T}}}\cdot \boldsymbol{G}\cdot \boldsymbol{V}-{{\boldsymbol{T}}^{\operatorname{T}}}\cdot \overline{\boldsymbol{\Pi }} \right].
\end{equation}

Substituting the previous expressions into equation (\ref{eqA6}) gives the energy balance equation
\begin{equation}\tag{A13}\label{eqA13}
\nabla \cdot \boldsymbol{p}=2\text{i}\omega ({{E}_{s}}+{{E}_{v}})-\omega \left( {{E}_{dv}}-{{E}_{ds}} \right).
\end{equation}

\subsection{Derivation of energy velocities and dissipation factors}\label{appendB}
This appendix derives the expressions of the energy velocity and the dissipation factor for an inhomogeneous plane wave propagating in attenuative anisotropic porous media.

From equation (\ref{eq20}), equation (\ref{eqA7}) can be recast as
\begin{equation}\tag{B1}\label{eqB1}
{{E}_{s}}=\frac{1}{4}\operatorname{Re}\left[ {{\boldsymbol{\Pi }}^{\operatorname{T}}}\cdot {{\boldsymbol{R}}^{\operatorname{T}}}\cdot \overline{\boldsymbol{\Pi }} \right].
\end{equation}

Using the equations (\ref{eq26}) and (\ref{eq37}), equation (\ref{eqB1}) becomes
\begin{equation}\tag{B2}\label{eqB2}
{{E}_{s}}=\frac{1}{4}\operatorname{Re}\left[ {{\overline{\boldsymbol{V}}}^{\operatorname{T}}}\cdot \overline{\boldsymbol{S}}\cdot \boldsymbol{R}\cdot {{\boldsymbol{S}}^{\text{T}}}\cdot \boldsymbol{V} \right],
\end{equation}
where the property of the matrix transpose has been used, that is, ${{\overline{\boldsymbol{V}}}^{\operatorname{T}}}\cdot \overline{\boldsymbol{S}}\cdot \boldsymbol{R}\cdot {{\boldsymbol{S}}^{\text{T}}}\cdot \boldsymbol{V}={{\boldsymbol{V}}^{\text{T}}}\cdot \boldsymbol{S}\cdot {{\boldsymbol{R}}^{\text{T}}}\cdot {{\overline{\boldsymbol{S}}}^{\operatorname{T}}}\cdot \overline{\boldsymbol{V}}$ due to ${{\boldsymbol{V}}^{\text{T}}}\cdot \boldsymbol{S}\cdot {{\boldsymbol{R}}^{\text{T}}}\cdot {{\overline{\boldsymbol{S}}}^{\operatorname{T}}}\cdot \overline{\boldsymbol{V}}$ is the first-rank matrix. 

With equations (\ref{eq37}) and (\ref{eq39}), we have
\begin{equation}\tag{B3}\label{eqB3}
\boldsymbol{G}\cdot \boldsymbol{V}=\boldsymbol{S}\cdot \boldsymbol{R}\cdot {{\boldsymbol{S}}^{\text{T}}}\cdot \boldsymbol{V}.
\end{equation}

Replacing equation (\ref{eqB3}) into equation (\ref{eqA8}) yields
\begin{equation}\tag{B4}\label{eqB4}
{{E}_{v}}=\frac{1}{4}\operatorname{Re}\left[ {{\overline{\boldsymbol{V}}}^{\operatorname{T}}}\cdot \boldsymbol{S}\cdot \boldsymbol{R}\cdot {{\boldsymbol{S}}^{\text{T}}}\cdot \boldsymbol{V} \right].
\end{equation}

The average stored energy density ${{E}_{a}}$ is identical to the sum of the average strain and kinetic energy densities, hence,
\begin{equation}\tag{B5}\label{eqB5}
\begin{split}
{{E}_{a}}=\, &{{E}_{s}}+{{E}_{v}}= \\
&\frac{1}{4}\operatorname{Re}\left[ {{\overline{\boldsymbol{V}}}^{\operatorname{T}}}\cdot \overline{\boldsymbol{S}}\cdot \boldsymbol{R}\cdot {{\boldsymbol{S}}^{\text{T}}}\cdot \boldsymbol{V}+{{\overline{\boldsymbol{V}}}^{\operatorname{T}}}\cdot \boldsymbol{S}\cdot \boldsymbol{R}\cdot {{\boldsymbol{S}}^{\text{T}}}\cdot \boldsymbol{V} \right].
\end{split}
\end{equation}

Using properties of complex numbers and dot product, equation (\ref{eqB5}) becomes
\begin{equation}\tag{B6}\label{eqB6}
{{E}_{a}}=\frac{1}{2}\operatorname{Re}\left[ {{\overline{\boldsymbol{V}}}^{\operatorname{T}}}\cdot \operatorname{Re}\left( \boldsymbol{S} \right)\cdot \boldsymbol{R}\cdot {{\boldsymbol{S}}^{\text{T}}}\cdot \boldsymbol{V} \right].
\end{equation}

Similarly, the average dissipated energy density (\ref{eqA12}) becomes
\begin{equation}\tag{B7}\label{eqB7}
{{E}_{d}}=\operatorname{Re}\left[ {{\overline{\boldsymbol{V}}}^{\operatorname{T}}}\cdot \operatorname{Im}\left( \boldsymbol{S} \right)\cdot \boldsymbol{R}\cdot {{\boldsymbol{S}}^{\text{T}}}\cdot \boldsymbol{V} \right].
\end{equation}

Using a convenient matrix notation, the average power flow density can be re-written as
\begin{equation}\tag{B8}\label{eqB8}
\operatorname{Re}\left[ \boldsymbol{p} \right]=-\frac{1}{2}\operatorname{Re}\left[ {{{\boldsymbol{\hat{e}}}}_{i}}{{({{\boldsymbol{\Theta }}_{i}}\cdot \boldsymbol{T})}^{\operatorname{T}}}\cdot \overline{\boldsymbol{V}} \right],
\end{equation}
where the matrices ${{\boldsymbol{\Theta }}_{i}}$, $i=1,2,3$ have the form
\begin{equation}\tag{B9.1}\label{eqB9.1}
{{\boldsymbol{\Theta }}_{1}}=\left[ \begin{matrix}
   1 & 0 & 0 & 0 & 0 & 0 & 0 & 0 & 0  \\
   0 & 0 & 0 & 0 & 0 & 1 & 0 & 0 & 0  \\
   0 & 0 & 0 & 0 & 1 & 0 & 0 & 0 & 0  \\
   0 & 0 & 0 & 0 & 0 & 0 & 1 & 0 & 0  \\
   0 & 0 & 0 & 0 & 0 & 0 & 0 & 0 & 0  \\
   0 & 0 & 0 & 0 & 0 & 0 & 0 & 0 & 0  \\
\end{matrix} \right],
\end{equation}
\begin{equation}\tag{B9.2}\label{eqB9.2}
{{\boldsymbol{\Theta }}_{2}}=\left[ \begin{matrix}
   0 & 0 & 0 & 0 & 0 & 1 & 0 & 0 & 0  \\
   0 & 1 & 0 & 0 & 0 & 0 & 0 & 0 & 0  \\
   0 & 0 & 0 & 1 & 0 & 0 & 0 & 0 & 0  \\
   0 & 0 & 0 & 0 & 0 & 0 & 0 & 0 & 0  \\
   0 & 0 & 0 & 0 & 0 & 0 & 0 & 1 & 0  \\
   0 & 0 & 0 & 0 & 0 & 0 & 0 & 0 & 0  \\
\end{matrix} \right],
\end{equation}
\begin{equation}\tag{B9.3}\label{eqB9.3}
{{\boldsymbol{\Theta }}_{3}}=\left[ \begin{matrix}
   0 & 0 & 0 & 0 & 1 & 0 & 0 & 0 & 0  \\
   0 & 0 & 0 & 1 & 0 & 0 & 0 & 0 & 0  \\
   0 & 0 & 1 & 0 & 0 & 0 & 0 & 0 & 0  \\
   0 & 0 & 0 & 0 & 0 & 0 & 0 & 0 & 0  \\
   0 & 0 & 0 & 0 & 0 & 0 & 0 & 0 & 0  \\
   0 & 0 & 0 & 0 & 0 & 0 & 0 & 0 & 1  \\
\end{matrix} \right].
\end{equation}

In terms of equations (\ref{eq29}) and (\ref{eq37}), equation (\ref{eqB8}) can be recast as
\begin{equation}\tag{B10}\label{eqB10}
\operatorname{Re}\left[ \boldsymbol{p} \right]=\frac{1}{2}\operatorname{Re}\left[ {{\overline{\boldsymbol{V}}}^{\operatorname{T}}}\left( {{{\boldsymbol{\hat{e}}}}_{i}}{{\boldsymbol{\Theta }}_{i}} \right)\cdot \boldsymbol{R}\cdot {{\boldsymbol{S}}^{\text{T}}}\cdot \boldsymbol{V} \right],
\end{equation}
where we use the fact that ${{\overline{\boldsymbol{V}}}^{\operatorname{T}}}\left( {{{\boldsymbol{\hat{e}}}}_{i}}{{\boldsymbol{\Theta }}_{i}} \right)\cdot \boldsymbol{R}\cdot {{\boldsymbol{S}}^{\text{T}}}\cdot \boldsymbol{V}={{\boldsymbol{V}}^{\operatorname{T}}}\cdot \boldsymbol{S}\cdot {{\boldsymbol{R}}^{\text{T}}}\cdot \left( {{{\boldsymbol{\hat{e}}}}_{i}}\boldsymbol{\Theta }_{i}^{\text{T}} \right)\cdot \overline{\boldsymbol{V}}$.

After substitution of the average power flow density and the average stored energy density, the energy velocity (\ref{eq43}) becomes
\begin{equation}\tag{B11}\label{eqB11}
{{\boldsymbol{V}}^{e}}=\frac{\operatorname{Re}\left[ {{\overline{\boldsymbol{V}}}^{\operatorname{T}}}\left( {{{\boldsymbol{\hat{e}}}}_{i}}{{\boldsymbol{\Theta }}_{i}} \right)\cdot \boldsymbol{R}\cdot {{\boldsymbol{S}}^{\text{T}}}\cdot \boldsymbol{V} \right]}{\operatorname{Re}\left[ {{\overline{\boldsymbol{V}}}^{\operatorname{T}}}\cdot \operatorname{Re}\left( \boldsymbol{S} \right)\cdot \boldsymbol{R}\cdot {{\boldsymbol{S}}^{\text{T}}}\cdot \boldsymbol{V} \right]}.
\end{equation}

Replacing equations (\ref{eqB2}) and (\ref{eqB7}) into equation (\ref{eq45}) yields the dissipation factor $Q_{1}^{-1}$ as
\begin{equation}\tag{B12}\label{eqB12}
Q_{1}^{-1}=\frac{{{E}_{d}}}{2{{E}_{s}}}=\frac{2\operatorname{Re}\left[ {{\overline{\boldsymbol{V}}}^{\operatorname{T}}}\cdot \operatorname{Im}\left( \boldsymbol{S} \right)\cdot \boldsymbol{R}\cdot {{\boldsymbol{S}}^{\text{T}}}\cdot \boldsymbol{V} \right]}{\operatorname{Re}\left[ {{\overline{\boldsymbol{V}}}^{\operatorname{T}}}\cdot \overline{\boldsymbol{S}}\cdot \boldsymbol{R}\cdot {{\boldsymbol{S}}^{\text{T}}}\cdot \boldsymbol{V} \right]}.
\end{equation}

The dissipation factor $Q_{2}^{-1}$ from equations (\ref{eq46}), (\ref{eqB6}), and (\ref{eqB7}) is
\begin{equation}\tag{B13}\label{eqB13}
Q_{2}^{-1}=\frac{{{E}_{d}}}{{{E}_{a}}}=\frac{2\operatorname{Re}\left[ {{\overline{\boldsymbol{V}}}^{\operatorname{T}}}\cdot \operatorname{Im}\left( \boldsymbol{S} \right)\cdot \boldsymbol{R}\cdot {{\boldsymbol{S}}^{\text{T}}}\cdot \boldsymbol{V} \right]}{\operatorname{Re}\left[ {{\overline{\boldsymbol{V}}}^{\operatorname{T}}}\cdot \operatorname{Re}\left( \boldsymbol{S} \right)\cdot \boldsymbol{R}\cdot {{\boldsymbol{S}}^{\text{T}}}\cdot \boldsymbol{V} \right]}.
\end{equation}

\subsection{Relationship between energy and phase velocities}\label{appendC}
This appendix verifies the relationship between the energy and phase velocities in attenuative anisotropic porous media.
Using equation (\ref{eq37}), equations (\ref{eqA1}) and (\ref{eqA2}) become
\begin{equation}\tag{C1}\label{eqC1}
-{{\boldsymbol{T}}^{\operatorname{T}}}\cdot {{\boldsymbol{S}}^{\operatorname{T}}}\cdot \overline{\boldsymbol{V}}={{\boldsymbol{T}}^{\operatorname{T}}}\cdot \overline{\boldsymbol{\Pi }},
\end{equation}
\begin{equation}\tag{C2}\label{eqC2}
-{{\overline{\boldsymbol{V}}}^{\operatorname{T}}}\cdot \boldsymbol{S}\cdot \boldsymbol{T}={{\overline{\boldsymbol{V}}}^{\operatorname{T}}}\cdot \boldsymbol{G}\cdot \boldsymbol{V}.
\end{equation}

Both equations (\ref{eqC1}) and (\ref{eqC2}) contain the complex Umov–Poynting vector (\ref{eqA5}). Therefore, according to equations (\ref{eq32}), (\ref{eq34}), and (\ref{eq35}), equations (\ref{eqC1}) and (\ref{eqC2}) can be simplified as
\begin{equation}\tag{C3}\label{eqC3}
2{{\overline{\boldsymbol{s}}}^{\text{T}}}\cdot \boldsymbol{p}={{\boldsymbol{T}}^{\operatorname{T}}}\cdot \overline{\boldsymbol{\Pi }}
\end{equation}
and
\begin{equation}\tag{C4}\label{eqC4}
2{{\boldsymbol{s}}^{\text{T}}}\cdot \boldsymbol{p}={{\overline{\boldsymbol{V}}}^{\operatorname{T}}}\cdot \boldsymbol{G}\cdot \boldsymbol{V},
\end{equation}
respectively.

Equation (\ref{eqC3}) plus equation (\ref{eqC4}) gives
\begin{equation}\tag{C5}\label{eqC5}
4\operatorname{Re}\left[ {{\boldsymbol{s}}^{\text{T}}} \right]\cdot \boldsymbol{p}={{\boldsymbol{T}}^{\operatorname{T}}}\cdot \overline{\boldsymbol{\Pi }}+{{\overline{\boldsymbol{V}}}^{\operatorname{T}}}\cdot \boldsymbol{G}\cdot \boldsymbol{V}.
\end{equation}

Replacing equation (\ref{eq31}) into equation (\ref{eqC5}), and the results are taken as the real part yields
\begin{equation}\tag{C6}\label{eqC6}
{{\boldsymbol{\hat{n}}}^{\text{T}}}\cdot \frac{\operatorname{Re}\left[ \boldsymbol{p} \right]}{{{E}_{a}}}=\frac{1}{\operatorname{Re}\left[ \tau  \right]},
\end{equation}
where equation (\ref{eqB5}) has been used.

Follow the expressions of the energy velocity (\ref{eq43}) and phase velocity (\ref{eq33}), we can get the relation as follows:
\begin{equation}\tag{C7}\label{eqC7}
{{\boldsymbol{\hat{n}}}^{\text{T}}}\cdot {{\boldsymbol{V}}^{e}}={{V}^{p}}.
\end{equation}
This equation implies that the relationship, that is, the phase velocity is the projection of the energy velocity in the propagation direction, still holds in attenuative anisotropic porous media.